\documentclass[%
 reprint,
superscriptaddress,
 amsmath,amssymb,
 aps,prl
]{revtex4-2}

\usepackage{graphicx}% Include figure files
\usepackage{dcolumn}% Align table columns on decimal point
\usepackage{bm}% bold math
\usepackage{minitoc}

\usepackage{tabularx}
\newcolumntype{b}{X}
\newcolumntype{s}{>{\hsize=.02\hsize}X}
\usepackage{multibib} % separate bibliographies for main text and supplementary
\newcites{S}{References for Supplementary Material}
\begin{document}
\doparttoc % Tell to minitoc to generate a toc for the parts
\faketableofcontents % Run a fake tableofcontents command for the partocs
\preprint{APS/123-QED}

\title{Coulomb Drag between a Carbon Nanotube and Monolayer Graphene}

\author{Laurel E. Anderson}
\affiliation{Department of Physics, Harvard University, Cambridge, Massachusetts 02138, USA}
\author{Austin Cheng}
\affiliation{Department of Applied Physics, Harvard University, Cambridge, Massachusetts 02138, USA}
\author{Takashi Taniguchi}
\affiliation{International Center for Materials Nanoarchitectonics, National Institute for Materials Science, Tsukuba 305-0044, Japan}
\author{Kenji Watanabe}
\affiliation{Research Center for Functional Materials, National Institute for Materials Science, Tsukuba 305-0044, Japan}
\author{Philip Kim}
\affiliation{Department of Physics, Harvard University, Cambridge, Massachusetts 02138, USA}
\affiliation{Department of Applied Physics, Harvard University, Cambridge, Massachusetts 02138, USA}
 \email{pkim@physics.harvard.edu}

\date{\today}

\begin{abstract}
We have measured Coulomb drag between an individual single-walled carbon nanotube (SWNT) as a one-dimensional (1D) conductor and the two-dimensional (2D) conductor monolayer graphene, separated by a few-atom-thick boron nitride layer. The graphene carrier density is tuned across the charge neutrality point (CNP) by a gate, while the SWNT remains degenerate. At high temperatures, the drag resistance changes sign across the CNP, as expected for momentum transfer from drive to drag layer, and exhibits layer exchange Onsager reciprocity. We find that layer reciprocity is broken near the graphene CNP at low temperatures due to nonlinear drag response associated with temperature dependent drag and thermoelectric effects. The drag resistance shows power-law dependences on temperature and carrier density characteristic of 1D Fermi liquid-2D Dirac fluid drag. The 2D drag signal at high temperatures decays with distance from the 1D source slower than expected for a diffusive current distribution, suggesting additional interaction effects in the graphene in the hydrodynamic transport regime.
\end{abstract}

\maketitle
\part{}
%\doparttoc

Coulomb interactions generate a plethora of novel emergent phenomena in condensed-matter systems, particularly when electronic confinement to fewer than three spatial dimensions increases the relative strength of potential to kinetic energy \cite{Lucas2018}. One experimental tool for studying interaction-driven effects in low-dimensional systems is Coulomb drag \cite{Narozhny2016,Gramila1991,Hill1996,Liu2017}. When a current is driven in a conductor that is near but electrically isolated from another, Coulomb interactions between the charge carriers in the two conductors generate a voltage drop in the “passive” conductor. The drag resistance is thus a direct probe of interlayer charge carrier interaction. Most of the past theoretical and experimental efforts have focused on drag between 2-dimensional (2D) conductors, such as electrons confined in semiconductor heterointerfaces \cite{Gramila1991,Hill1996} and graphene \cite{Liu2017,Kim2011,Gorbachev2012}, revealing several new emergent phenomena including exciton condensation under strong magnetic fields \cite{Liu2017}. Drag experiments have also been performed between 1-dimensional (1D) conductors \cite{Yamamoto2006,Laroche2011,Laroche2014}, showing signatures of Wigner crystal and Luttinger liquid behavior. 

Coulomb drag experiments between mixed-dimensional systems, e.g. 1D-2D conductors, have also been conceived \cite{Sirenko1992,Lyo2003} to investigate the effects of dimensionality on electron-electron interactions. Such a system was recently probed experimentally \cite{Mitra2020a} using an InAs nanowire as 1D conductor and graphene as 2D conducting layer. This recent 1D-2D drag experiment shows an anomalous temperature and density dependent drag response that might be related to energy drag \cite{Song2012,Song2013,Song2013a} due to the large mismatch in thermal conductivities between InAs and graphene. However, the breakdown of layer (Onsager) reciprocity and subsequent thermopower measurements in these devices \cite{Mitra2020} suggest local heating-induced thermoelectric effects may also play a substantial role in the reported drag results.

In this letter, we report Coulomb drag in a new 1D-2D conducting system, a metallic single-walled carbon nanotube (SWNT) and monolayer graphene separated by an atomically thin (2-4 nm) insulating barrier of hexagonal boron nitride (hBN). Since SWNTs and graphene have similar linear dispersion relations with comparable Fermi energies and work functions \cite{Laird2015,CastroNeto2009}, the interaction-driven momentum and energy transfer between carriers in separate layers are enhanced, amplifying the drag signal. Due to the small ($\sim$2 nm) diameter of the SWNT, driving current in the nanotube provides an extremely localized 1D drag source in the graphene channel. We measure density, temperature, and distance dependence of the drag effect to probe the carrier interactions between these 1D and 2D conductors. This work is an early step toward experimentally addressing the specific impact of increased confinement on interaction effects, and may also be a new test system for hydrodynamics in graphene.

A scanning electron microscope image of an example SWNT-graphene drag device is shown in Figure \ref{fig1}(a). Details of the fabrication are given in the Supplemental Material (SM) \cite{SM}. In brief, monolayer graphene is encapsulated in hBN and then transferred on top of a metallic SWNT. The hBN flake separating the SWNT and graphene is 2-4 nm thick, so that the two conductors are sufficiently close together for interlayer Coulomb interactions, but they remain electrically isolated, without a significant tunneling current. The graphene and SWNT have individual electrical contacts, allowing them to be characterized separately. We can use graphene or nanotube as either drive or drag layer. While we focus on one SWNT in the following discussion, several SWNT/graphene devices were measured, and similar results were obtained (see SM \cite{SM}).

\begin{figure}
\includegraphics[width=0.49\textwidth]{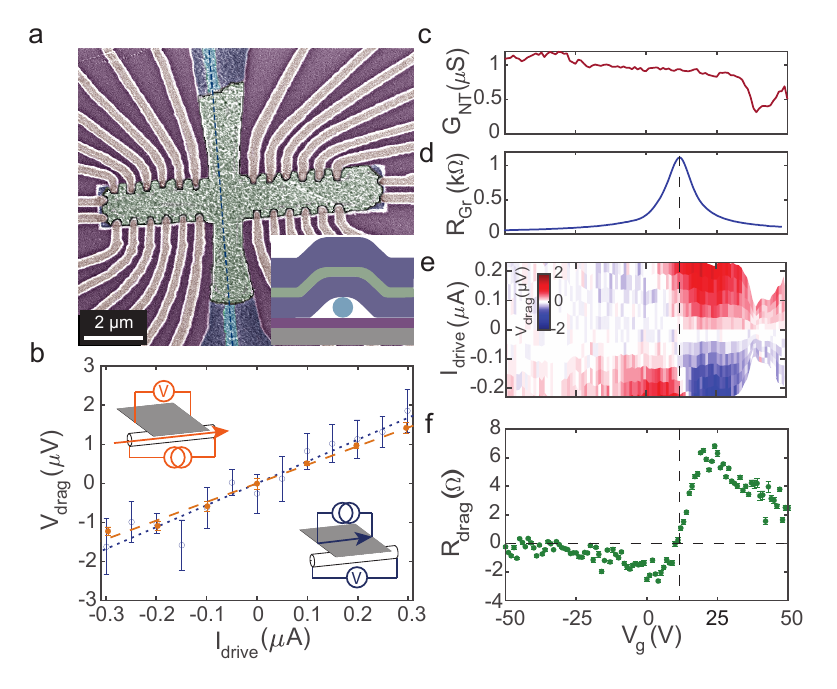}
\caption{\label{fig1}
(a) False color scanning electron microscope image of a typical SWNT-graphene drag device. Graphene (green) is encapsulated in hBN (dark blue) and transferred on top of a metallic SWNT (dashed line in center of blue charged region). Electrical contacts (gold) are made to the graphene and SWNT. Inset: cross-section schematic of the device. (b) $V_{drag}$ versus $I_{drive}$ for reciprocal layer configurations: nanotube-drive, graphene-drag (orange, filled symbols) and graphene-drive, nanotube-drag (blue, open symbols). Data were taken at $T = 300 \; \textrm{K}$ and $V_g = 21 \; \textrm{V}$, with averaging gate voltage window $\Delta V=\pm 1$ V to enhance the signal-to-noise ratio. Dashed curves are lines of best fit. Additional details can be found in the SM \cite{SM}. (c) SWNT conductance as a function of gate voltage. The dip is a local conductance minimum, not the charge neutrality point (see SM \cite{SM} for discussion). (d) Graphene resistance as a function of gate voltage. (e) $V_{drag}$ in graphene versus SWNT $I_{drive}$ and $V_g$. (f) $R_{drag}$ versus $V_g$. Dashed line marks zero drag signal.}
\end{figure}

Measurements of the drag resistance were performed by applying DC current $I_{drive}$ through the drive layer (SWNT or graphene) while the voltage $V_{drag}$ was measured in the drag layer (graphene or SWNT). Example data for both configurations are presented in Figure \ref{fig1}(b). When using the SWNT as drive layer, $V_{drag}$ in graphene is measured with the voltage probes nearest the SWNT, at a distance $x=800$ nm away (closed circles). At temperature $T = 300 \; \textrm{K}$, there is a linear relationship between $I_{drive}$ and $V_{drag}$, whether the SWNT or the Gr is the drive layer. Using the graphene as the drive layer and measuring $V_{drag}$ across the SWNT (open circles) results in a noisier signal than the reciprocal drag scheme, due to the higher resistance of the SWNTs (see SM \cite{SM} for discussion). Even so, both biasing configurations yield the same current-voltage relationship. This Onsager reciprocity when drive and drag layers are exchanged demonstrates that the system is in the linear response regime, allowing the extraction of the drag resistance from the slope: $R_{drag} = \Delta V_{drag}/\Delta I_{drive}$.

In our devices, the SWNTs are beneath the graphene (Fig. \ref{fig1}(a) inset), enabling the carrier densities in both SWNT and graphene to be tuned by a voltage $V_g$ applied to a back gate. Figure \ref{fig1}(c) and (d) show the conductance $G_{NT}$ of SWNT and resistance $R_{Gr}$ of graphene, respectively, as a function of $V_g$ measured at $T = 300\; \textrm{K}$. The gradual decrease of $G_{NT}$ as $V_g$ increases indicates the SWNT is hole-doped. In the graphene, $R_{Gr}$ exhibits a peak corresponding to the charge neutrality point (CNP) around gate voltage $V_0=13 \; \textrm{V}$. We also measure the drag response as a function of $V_g$, as shown in Figure \ref{fig1}(e). We extract $R_{drag}$ in the linear response regime in as a function of $V_g$, as described above. For $V_g < V_0$, $V_{drag}$ and $I_{drive}$ have opposite sign, while for $V_g > V_0$, they have the same sign. As shown in Figure \ref{fig1}(f), $R_{drag}$ thus changes sign at $V_g = V_0$ where the dominant carrier type in graphene switches from electrons to holes. This behavior is qualitatively similar to previous measurements of momentum-transfer Coulomb drag in double-layer graphene systems \cite{Kim2011,Gorbachev2012}. The higher magnitude of e-h compared to h-h drag can be attributed to the higher density of holes in the SWNT at more negative gate voltages. Due to heavy SWNT doping, the $k_F$’s of the SWNT and graphene do not overlap within our experimental gate window (further discussion in SM \cite{SM}), preventing us from investigating the double neutrality point, where the chiral nature of the 1D-2D Dirac system can be explored \cite{Hwang2011}.

\begin{figure}
\includegraphics[width=0.49\textwidth]{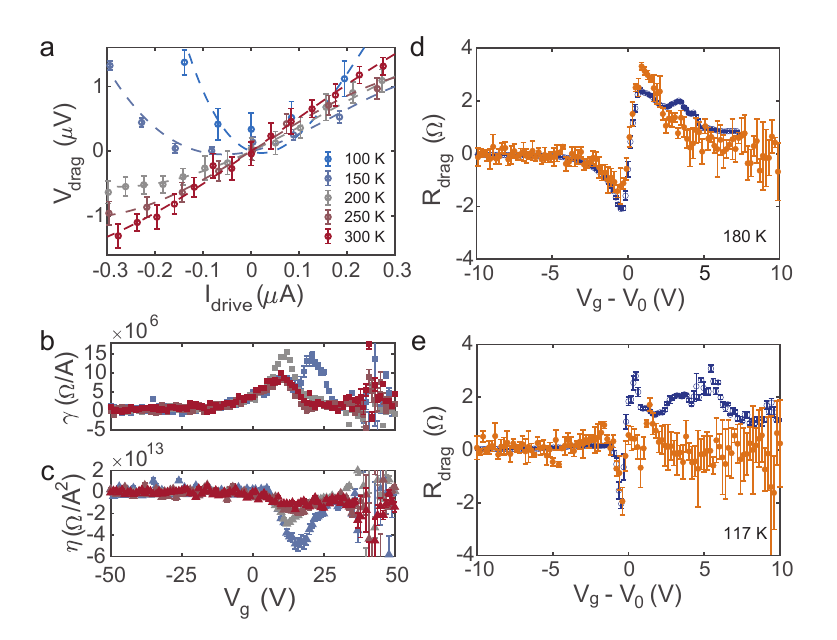}
\caption{\label{fig2} (a) $V_{drag}$ versus $I_{drive}$ at $V_g = 22$ V for varying temperatures. Dashed lines are 3rd-order polynomial fits. (b) 2nd-order coefficient versus $V_g$. (c) 3rd-order coefficient versus $V_g$. Regions of noisy data in (b) and (c) are due to SWNT local conductance dip limiting $I_{drive}$. (d) $R_{drag}$ from linear fit versus $V_g$ (subtracting the graphene CNP voltage) for reciprocal configurations at $T = 180\; \textrm{K}$ (same colors and symbols as Fig. \ref{fig1}(b)). (e) Same measurement at $T = 117\; \textrm{K}$.}
\end{figure}

As T decreases, the relationship between drive current and drag voltage becomes increasingly nonlinear (Fig. \ref{fig2}(a)). To quantitatively address this change, we fit $V_{drag}$ with a 3rd-order polynomial in $I_{drive}$: $V_{drag}=I_{drive} R_{drag}+\gamma I_{drive}^2+\eta I_{drive}^3$, where $\gamma$ and $\eta$ are fitting coefficients. The nonlinear effect sensitively varies with $V_g$; Figure \ref{fig2}(b-c) shows the $V_g$ dependence of $\gamma$ and $\eta$  at several fixed temperatures. We find that $\gamma>0$ and $\eta<0$ for all gate voltage ranges we probe, and both quantities have larger magnitude nearer the CNP of graphene ($V_g \approx V_0$) and at lower temperatures. This increasingly nonlinear effect also breaks Onsager layer reciprocity at low temperatures. As shown in Figure \ref{fig2}(d) and (e), the drag resistance from SWNT-drive and graphene-drive configurations show progressively worse correspondence at lower temperatures, as the nonlinear part of the relation between $I_{drive}$ and $V_{drag}$ becomes appreciable.

Higher-order dependence of $V_{drag}$ on $I_{drive}$ is best explained by development of a temperature gradient in the SWNT due to the Peltier effect or Joule heating, which can be efficiently transferred to the nearby graphene \cite{Song2012,Song2013,Song2013a}. Such a temperature gradient in the graphene generates a thermoelectric voltage and causes a temperature-dependent change in $R_{drag}$. Both can give rise to nonlinear terms in $V_{drag}$ (see SM \cite{SM} for further analysis).

\begin{figure}
\includegraphics[width=0.49\textwidth]{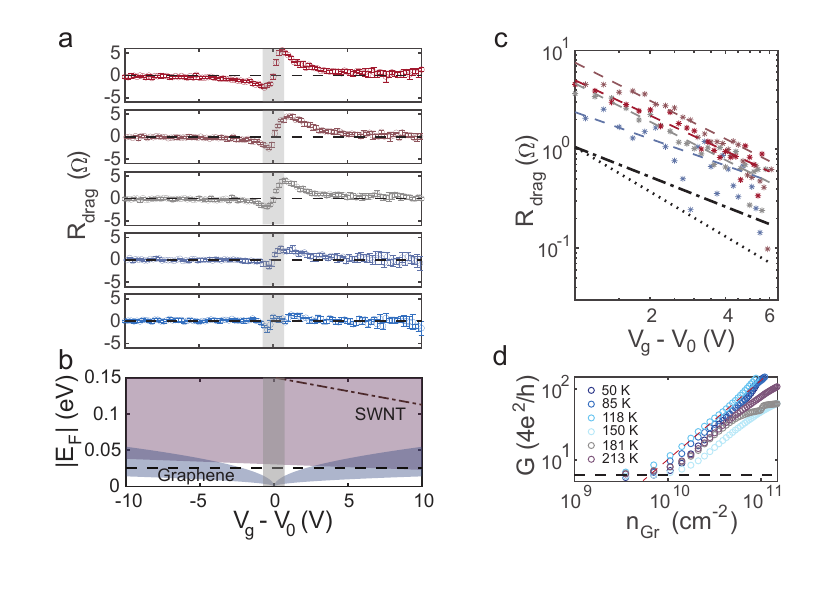}
\caption{\label{fig3}(a) Drag resistance as a function of $V_g$ at various temperatures: 265 K (top), 235 K, 200 K, 160 K, 130 K (bottom). Disorder-dominated range is indicated by gray shading. b) Estimated range of Fermi energies for SWNT (purple) and graphene (blue) in a range of $V_g$ near the graphene CNP. Dot-dashed line is approximate center of SWNT $E_F$ range. See SM \cite{SM} for details of calculation. (c) Log-log plot of $R_{drag}$ versus $V_g-V_0$ at selected temperatures, with $R_{drag}\propto (V_g-V_0)^{-1}$ (dot-dashed) and $R_{drag}\propto (V_g-V_0)^{-1.5}$ (dotted) for comparison. (d) Graphene conductance versus charge carrier density for temperatures in (a). The residual carrier density $\delta n$ is estimated by the intersection of the line at minimum conductivity (black) and a linear fit to log(G) away from charge neutrality (dark red dashed line shows example fit for $T = 118 \; \textrm{K}$).}
\end{figure}

To avoid the nonlinear drag phenomena discussed above, we focus on linear drag resistance measured at small drive current at relatively high temperature ($T > 100 \; \textrm{K}$). Figure \ref{fig3}(a) shows $R_{drag}$ as a function of gate voltage $V_g$, referenced to the CNP of graphene $V_0$, at different fixed temperatures in this regime. In general, $R_{drag}$ changes sign at the graphene CNP, and that $|R_{drag}|$ grows linearly, peaks, then rapidly decreases as the graphene carrier density, $n_{Gr}\propto V_g-V_0$, increases. Fig. \ref{fig3}(b) shows that $R_{drag}\sim (V_g-V_0)^{-\beta}$, where $1<\beta <2$ at different temperatures. This behavior resembles 2D-2D graphene drag, where $1<\beta <2$ has also been observed \cite{Kim2011,Gorbachev2012}.

To determine the carrier densities (and thus Fermi energies) of each conductor as a function of $V_g$, we employ a finite element analysis of the graphene channel and SWNT together with the hBN separation layers and silicon back gate (Fig. \ref{fig1}(a) inset). Since the SWNT locally screens the graphene channel from the back gate, the local carrier density in graphene is reduced in the graphene channel directly above the SWNT and maximized away from the SWNT. To estimate the carrier density (and thus chemical potential) of the SWNT, we also need to consider device geometry and quantum capacitance (detailed discussion in \cite{SM}). Figure \ref{fig3}(b) summarizes the estimated upper and lower bounds of the Fermi energies of graphene $E_F^{Gr}$ and SWNT $E_F^{NT}$. While the SWNT remains a heavily p-doped degenerate 1D conductor in the experimental gate voltage range, our analysis suggests that $E_F^{Gr}$ is comparable to or even smaller than $k_BT$ in the temperature range $T > 100 \; \textrm{K}$, where $k_B$ is the Boltzmann constant, for all $V_g$ where the drag signal is measurable.

Near the CNP of the graphene channel, disorder becomes more relevant, creating charge puddles \cite{Martin2008}. The $n_{Gr}$-dependent conductance of the graphene channel is accordingly expected to saturate at low temperatures for $|n_{Gr}|<\delta n$, where $\delta n$ is the residual density due to charge puddles, which can be estimated from the temperature-dependent conductance $G$ of the graphene \cite{Couto2014}. Figure \ref{fig3}(d) shows $G(n_{Gr})$ measured in the graphene channel of our device for $T \lesssim 150 \; \textrm{K}$. From the saturation of $G(n_{Gr})$ near the CNP, we estimate $\delta n\approx 1.1 \times 10^{10}\; \textrm{cm}^{-2}$.  For $|n_{Gr}|<\delta n$, the electron and hole contributions of Coulomb drag cancel, resulting in linearly vanishing $R_{drag}$ with $n_{Gr}$ as observed in the experiment (shaded region in Fig. \ref{fig3}(a)). We also estimate the puddle energy scale $k_B T_d=\hbar v_F \sqrt{\pi \delta n}$ where $\hbar$ and $v_F=10^6 \;\textrm{m/s}$ are the reduced Plank constant and Fermi velocity of graphene, respectively. From $\delta n$ experimentally obtained above, we find the disorder temperature scale $T_d\approx 140 \; \textrm{K}$, which separates the low temperature regime where disorder effects are dominant and the high temperature regime where thermal broadening is appreciable.

\begin{figure}
\includegraphics[width=0.49\textwidth]{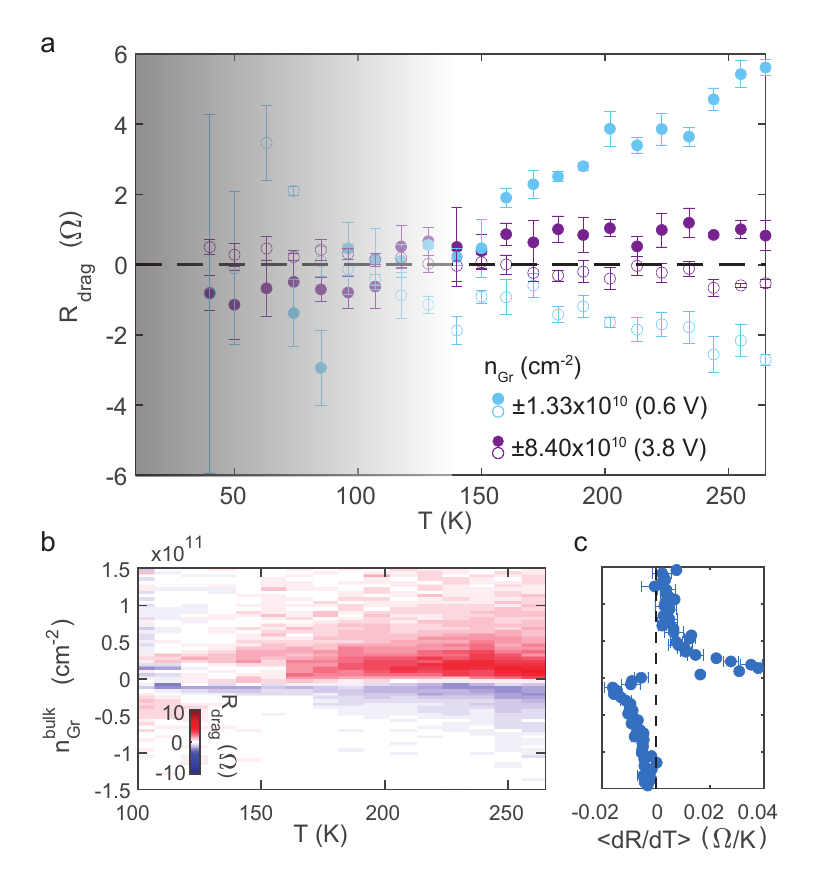}
\caption{\label{fig4} (a) Drag resistance as a function of temperature for various graphene charge carrier densities (stated value is the upper bound). For $T < 140 \; \textrm{K}$ (the shaded region in the graph), nonlinear effects increasingly dominate the signal.(b) Drag resistance versus temperature and graphene carrier density. (c) Average $\frac{dR_{drag}}{dT}$ for $T>T_d$ versus temperature at a range of graphene carrier densities.}
\end{figure}

We find the drag near the graphene CNP depends sensitively on temperature. Figure \ref{fig4}(a) shows $T$-dependent $R_{drag}$ at fixed density (reported as $n_{Gr}^{bulk}$, the upper bound of the estimated graphene carrier density). For $n_{Gr}^{bulk}=\pm 1.3 \times 10^{10} \; \textrm{cm}^{-2}$, close to the peak value of $|R_{drag}(n_{Gr}^{bulk})|$, $R_{drag}$ increases linearly in $T-T_d$ in the high temperature regime ($T>T_d$). In the low temperature regime ($T<T_d$), however, the linear response $R_{drag}$ is difficult to determine, due to the nonlinear drag effects and broken Onsager reciprocity discussed above. At larger density (e.g. $n_{Gr}^{bulk}=\pm 8.4 \times 10^{10} \; \textrm{cm}^{-2}$, far from the CNP), we observe a similar trend, although $|R_{drag}|$ is reduced. A broader range of the density and temperature dependent $R_{drag}(n_{Gr}^{bulk},T)$ is shown in Figure \ref{fig4}(b), where the magnitude of the drag resistance appears to increase approximately linearly at all densities. For $T > T_d$, the density dependence of $\frac{dR_{drag}}{dT}$ behaves similarly to $R_{drag}(n_{Gr}^{bulk})$ [Fig. \ref{fig4}(c)].

The temperature dependent drag behavior discussed above is distinctly different from 2D-2D drag in graphene, where a crossover between  $R_{drag}(T) \sim$ constant and $R_{drag}(T) \sim T^{-2}$ is expected \cite{Narozhny2012} in the parameter range of our experimental regime, $E_F^{Gr}\lesssim k_B T\sim E_F^{NT}$. For 1D-2D drag between SWNT and graphene, Badalyan and Jauho calculated the Coulomb drag effect in the Fermi liquid regime of both conductors ($k_B T \ll E_F^{NT},E_F^{Gr}$) \cite{Badalyan2020}, predicting $R_{drag}(T)\sim T^{\alpha}$, where $\alpha \approx 3.7$ at low temperatures. A more general theory of 1D-2D drag \cite{Lyo2003} predicts a transition to $1<\alpha <2$ at higher temperatures ($T>T_d$). While a more extensive model extending to the nondegenerate Dirac fluid limit in the presence of disorder is required for further quantitative comparison, our experiments show qualitatively similar behavior in the high temperature limit. 

Finally, we studied the relationship between the drag signal strength and the distance of the graphene voltage probes from the SWNT. Previous experiments have demonstrated signatures of hydrodynamic electron flow from current injection into a rectangular graphene channel \cite{Bandurin2016,Berdyugin2019,Ku2020}, with discernable effects even at room temperature \cite{Berdyugin2019,Ku2020}. Viscosity of the electron fluid causes the injected current to draw neighboring regions along with it, resulting in a negative potential near the injection contacts and creating current whirlpools in certain confined geometries \cite{Bandurin2016,Berdyugin2019,Levitov2016,Pellegrino2017}. Our SWNT-graphene Coulomb drag device geometry provides a unique experimental probe of hydrodynamic flow of graphene charge carriers, as the current flowing in the SWNT generates a direct dragging force on the graphene carriers without injecting current in graphene. This approach should have the benefit of eliminating diffusive “spray” from the contacts that could mask hydrodynamic transport signatures.

\begin{figure}
\includegraphics[width=0.49\textwidth]{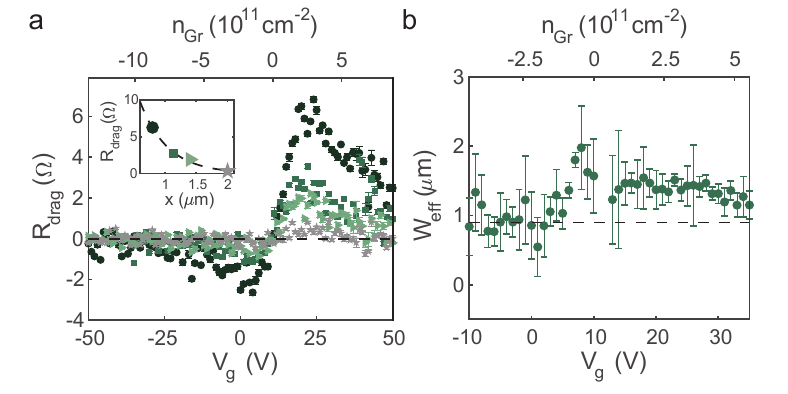}
\caption{\label{fig5} (a) $R_{drag}$ versus $V_g$ at $T = 300 \; \textrm{K}$ for pairs of voltage probes at increasing distance $x$ from the SWNT: 800 nm (circles), 1.2 $\mu$m (squares), 1.4 $\mu$m (triangles), and 2 $\mu$m (stars). Inset: $R_{drag}$ at  $V_g= 24 \; \textrm{V}$ for increasing distances. Dashed curve is an exponential fit (see main text). (b) Effective channel width $W_{\textrm{eff}}$ versus $V_g$, extracted from fit for $R_{drag}$ at $T = 300 \; \textrm{K}$. Dashed line marks actual device width ($W = 1 \; \mu$m).}
\end{figure}

Figure \ref{fig5}(a) shows $R_{drag}$ measured at pairs of voltage probes in the graphene channel laterally displaced by distance $x$ away from the SWNT. $R_{drag}$ decreases as $x$ increases, becoming almost unmeasurable for $x > 2\; \mu$m. In Ohmic transport, such a diminishing drag signal can be understood with a diffusive model, where the escaping current density in the graphene just above the SWNT is expected to decay as $J_{esc}(x)\sim e^{-\pi x/W}$, where $W$ is the channel width \cite{Abanin2011}. In the diffusive transport regime, we therefore expect driving current in the SWNT to cause a drag voltage in the probes at distance $x$ away following $R_{drag}(x)\sim e^{-\pi x/W}$. The inset of Figure \ref{fig5}(a) shows that the measured $R_{drag}(x)$ follows such an exponential decay. We obtain the effective channel width $W_{\textrm{eff}}$ by fitting this functional dependence. Figure 5(b) shows $W_{\textrm{eff}}$ as a function of $V_g$. Interestingly, we find $W_{\textrm{eff}}$ is larger than the physical channel width $W = 1 \; \mu$m in our device when the graphene is in the Dirac fluid regime, enhanced by about a factor of 2 at the CNP. Based on previous observations that the electron fluid of graphene is highly viscous in this temperature range near the CNP \cite{Bandurin2016,Bandurin2018}, the increase in $W_{\textrm{eff}}$ may hint at a hydrodynamic contribution to the transport behavior.

In summary, we present an experimental study of mixed-dimensional Coulomb drag between a SWNT and graphene. Our drag measurements in a SWNT-graphene heterostructure are qualitatively consistent with momentum transfer between the drive and drag layers, although we also observe an onset of nonlinearity due to local energy transfer combined with temperature dependent drag effects at lower temperatures. Within the linear response regime, the dependences on temperature, carrier density, and distance have subtleties that suggest an interplay of different mechanisms at work in this novel hybrid system. Further measurements with higher spatial resolution, such as current imaging \cite{Sulpizio2019,Ku2020}, would be necessary to gain a deeper understanding of the current flow patterns, and samples with less disorder should amplify hydrodynamic transport signatures in the graphene \cite{Levitov2016}.

\begin{acknowledgments}
We acknowledge Leonid Levitov, Patrick Ledwith, and Artem Talanov for useful discussions. Sample preparation and device fabrication was supported by ONR MURI (N00014-16-1-2921). P.K. and A.C. acknowledge support from the ARO (W911NF-17-1-0574) for a part of measurement and analysis. L.E.A. acknowledges support from the ARO through the NDSEG Fellowship Program. K.W. and T.T. acknowledge support from the Elemental Strategy Initiative conducted by the MEXT, Japan (grant number JPMXP0112101001) and JSPS KAKENHI (grant numbers 19H05790 and JP20H00354). This work was performed, in part, at the Center for Nanoscale Systems (CNS), a member of the National Nanotechnology Infrastructure Network, which is supported by the NSF under Grant No. ECS-0335765. CNS is part of Harvard University.
\end{acknowledgments}
%\bibliographystyle{apsrev4-2}
%\bibliography{CNT-Gr-main-v3}% Produces the bibliography via BibTeX.


%%%%%%%%%% Merge with supplemental materials %%%%%%%%%%
\clearpage
\pagebreak
\widetext
\begin{center}
\textbf{\Large Supplemental Material for ``Coulomb Drag Between a Carbon Nanotube and Monolayer Graphene''}
\end{center}
%%%%%%%%%% Merge with supplemental materials %%%%%%%%%%
%%%%%%%%%% Prefix an "S" to all equations, figures, tables and reset the counter %%%%%%%%%%
\setcounter{equation}{0}
\setcounter{figure}{0}
\setcounter{table}{0}
\setcounter{page}{1}
\makeatletter
\renewcommand{\theequation}{S\arabic{equation}}
\renewcommand{\thefigure}{S\arabic{figure}}
\renewcommand{\bibnumfmt}[1]{[S#1]}
\renewcommand{\citenumfont}[1]{S#1}
%%%%%%%%%% Prefix an "S" to all equations, figures, tables and reset the counter %%%%%%%%%%
%\part{}
%\mtcsetrules{parttoc}{off}
%\parttoc

\textbf{\large Contents}\\
\\
\begin{tabularx}{0.9\textwidth}{bs}
\hspace{5mm}I: Fabrication of SWNT-graphene devices & \multicolumn{1}{r}{\pageref{sec1}} \\
\hspace{10mm}A: SWNT synthesis, characterization, and transfer & \multicolumn{1}{r}{\pageref{sec1A}} \\
\hspace{10mm}B: Graphene-hBN heterostructure fabrication and transfer & \multicolumn{1}{r}{\pageref{sec1B}} \\
\hspace{10mm}C: Additional nanofabrication & \multicolumn{1}{r}{\pageref{sec1C}} \\
\hspace{5mm}II: SWNT conductance & \multicolumn{1}{r}{\pageref{sec2a}} \\
\hspace{5mm}III: DC measurement technique & \multicolumn{1}{r}{\pageref{sec2}} \\
\hspace{5mm}IV: Additional Onsager data & \multicolumn{1}{r}{\pageref{sec2b}} \\
\hspace{5mm}V: Graphene chemical potential and its effect on Coulomb drag  & \multicolumn{1}{r}{\pageref{sec3}} \\
\hspace{5mm}VI: Estimation of SWNT and graphene carrier densities & \multicolumn{1}{r}{\pageref{sec4}} \\
\hspace{5mm}VII: Derivation of nonlinear contributions to the drag voltage & \multicolumn{1}{r}{\pageref{secT}} \\
\hspace{5mm}VIII: Comparison of individual layer transport and drag resistance in different measurements & \multicolumn{1}{r}{\pageref{sec5}} \\
\hspace{5mm}References & \multicolumn{1}{r}{\pageref{Sbib}}
\end{tabularx}

\section{\label{sec1}I: Fabrication of SWNT-graphene devices}
More complete details of some device fabrication processes are described elsewhere, particularly in \cite{S_Cheng2019}.
\subsection{\label{sec1A}A: SWNT synthesis, characterization, and transfer}
\begin{figure}
    \centering
    \includegraphics[width=0.9\textwidth]{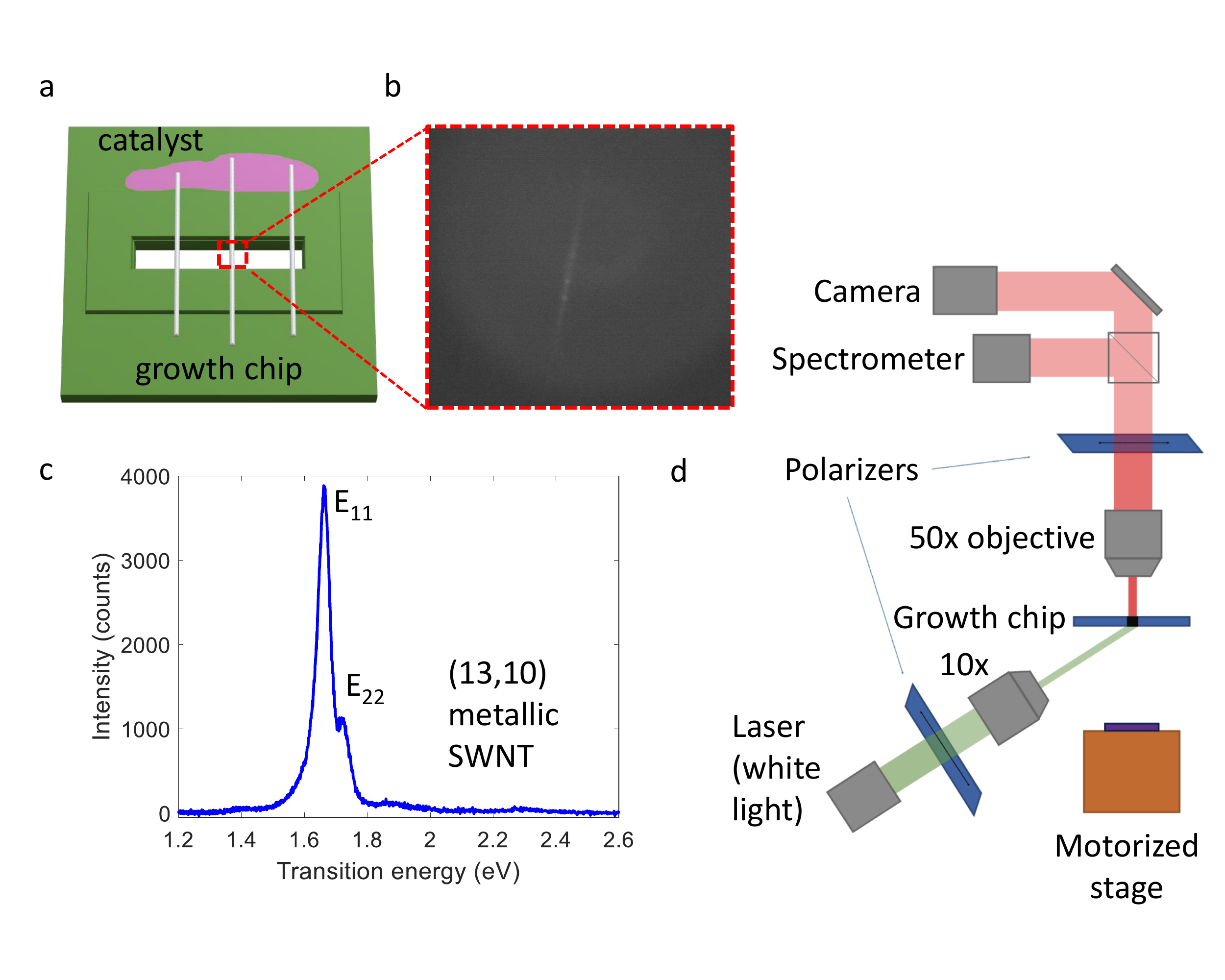}
    \caption{Nanotube suspended growth, imaging and characterization. (a) Schematic of nanotube growth chip with catalyst and central slit. (b) Infrared image of suspended SWNT. (c) Optical excitation spectrum of an example metallic SWNT. (d) Schematic of Rayleigh spectroscopy, imaging and transfer stage setup.}
    \label{fig:S1}
\end{figure}

Carbon nanotubes are grown in a chemical vapor deposition (CVD) furnace using the method described in \cite{S_Sfeir2006}. The growth substrate is a 5 mm $\times$ 5 mm silicon chip with a slit in the center [Fig. \ref{fig:S1}(a)], oriented perpendicular to the gas flow direction. A cobalt-molybdenum-based catalyst is applied to the chip on the side of the slit nearer to the gas inlet, so that nanotubes grow suspended across the slit. Suspended nanotubes were located and characterized using Rayleigh scattering spectroscopy \cite{S_Sfeir2006} and imaging [Fig. \ref{fig:S1}(d)]. By matching peaks in Rayleigh scattering intensity with nanotube optical transition energies, the chiral indices (and thus diameter and metallic/semiconducting nature) of the nanotube can be determined [Fig. \ref{fig:S1}(c)]. The scattered light is also routed to a camera, providing an image of the nanotube spanning the slit [Fig. \ref{fig:S1}(b)]. All the SWNTs used in devices described in this paper were metallic.

\begin{figure}
    \centering
    \includegraphics[width=0.9\textwidth]{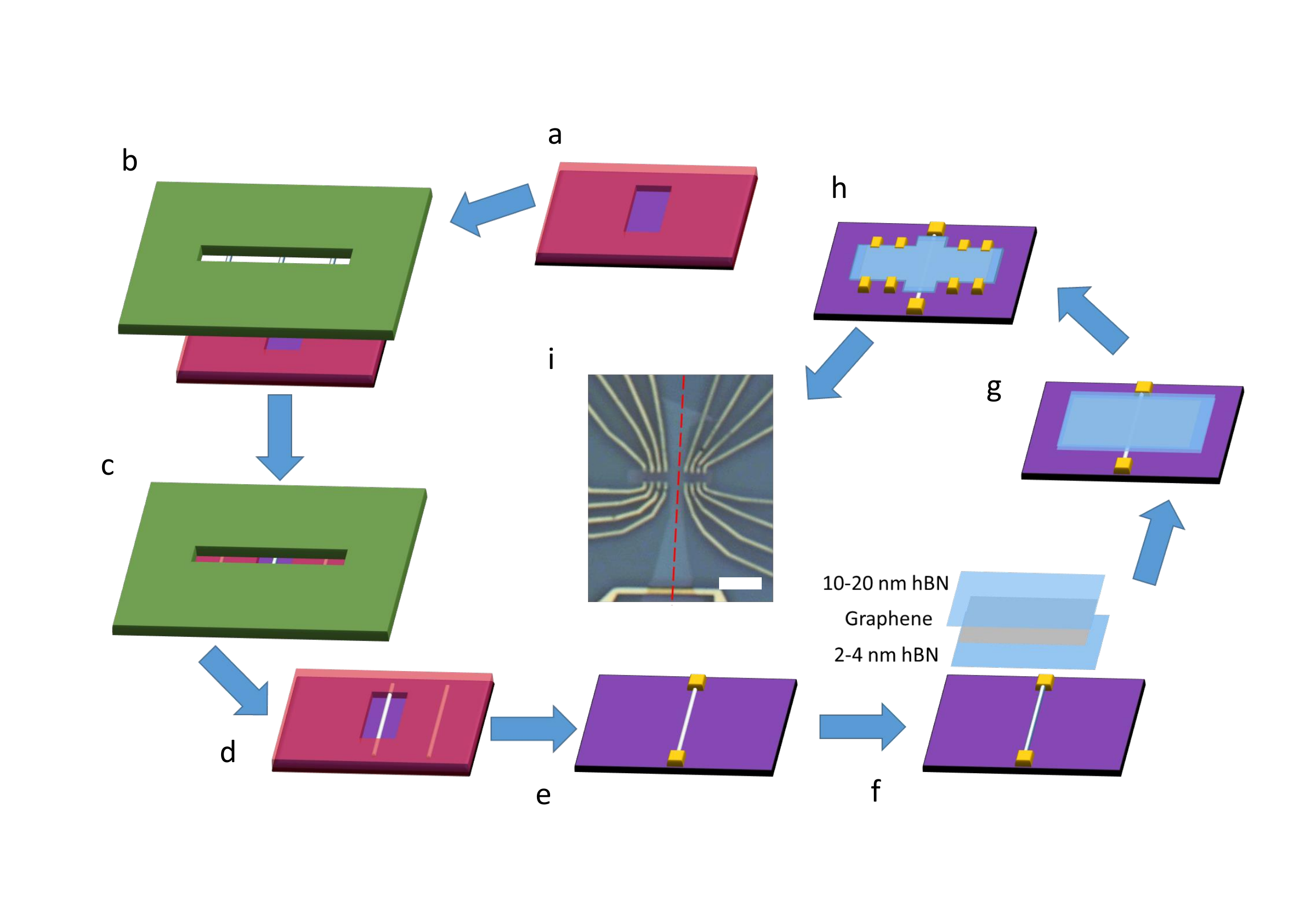}
    \caption{Fabrication sequence for SWNT-graphene devices. (a) PMMA window for SWNT transfer. (b) Alignment of growth chip with CNTs to window. (c) Mechanical contact between growth chip and target chip. (d) Growth chip removal; SWNTs are captured by PMMA. (e) PMMA removal and patterning/deposition of SWNT contacts. (f-g) Transfer of hBN and graphene flakes to contacted SWNT. (h) Heterostructure shaped by etching, patterning/deposition of graphene contacts. (i) Optical microscope image of a final device, with SWNT position indicated by red dashed line. Gold electrode at bottom of image is one of the SWNT contacts. Scale bar is 2 $\mu$m.}
    \label{fig:S2}
\end{figure}

Electron beam (e-beam) lithography was used to define a $\sim 50 \times 50 \; \mu$m resist-free window on a SiO$_2$/Si chip coated with $\sim$100 nm of 495 K Polymethyl methacrylate (PMMA) A4 resist. The resist layer helps the SWNTs transfer to the chip \citeS{S_Huang2005}. When a single SWNT with the desired characteristics has been located, it is aligned with the heterostructure so that it crosses the center of the PMMA window [Fig. \ref{fig:S2}(a-b)]. The growth chip and PMMA-coated sample are pressed together until mechanical contact is evident [Fig. \ref{fig:S2}(c)], then heated to 180 °C for 5 minutes to soften the resist. The chips are then cooled to 90 °C and slowly separated [Fig. \ref{fig:S2}(d)]. Successful SWNT transfer is confirmed by scanning electron microscope (SEM) or atomic force microscope (AFM) imaging. The same PMMA window can be reused several times in the event of unsuccessful transfer, or to transfer multiple SWNTs for parallel device fabrication. Finally, the PMMA is removed by high-temperature annealing in vacuum.

To anchor the SWNT to the substrate and confirm its suitability to be incorporated into a device, electrical contacts were made to either end of a 30-50 $\mu$m section of the SWNT, inside the window region [Fig. \ref{fig:S2}(e)]. The mask for the contacts was defined by e-beam lithography, and metal was deposited by e-beam evaporation (10 nm Cr/60 nm Au). Many other recipes for high-quality SWNT contacts have been reported in literature, including some measurements with quantum-limited contact resistance \citeS{S_Cao2015,S_Huang2015,S_Kim2005,S_Pitner2019}. However, this recipe was found to remain the most reliable through the additional fabrication steps.
\subsection{\label{sec1B}B: Graphene-hBN heterostructure fabrication and transfer}
Boron nitride-encapsulated graphene heterostructures were prepared using standard techniques \citeS{S_Dean2010,S_Wang2013}. The top hBN flake (20-40 nm) is picked up with a polypropylene carbonate (PPC) film on a Polydimethylsiloxane (PDMS) stamp, and then used to pick up graphene and bottom hBN (2-5 nm) flakes. It was critical to ensure that the bottom hBN flake was both thin and large enough to completely cover the graphene (at least in the region intended to be near the SWNT), in order to allow interaction between the graphene and SWNT without shorting. Once assembled on a stamp, the stack was transferred on top of a contacted SWNT [Fig. \ref{fig:S2}(f-g)]. The PPC and stack were detached from the stamp by heating to 150° C to melt the PPC, which was then removed by high-temperature vacuum annealing.
\subsection{\label{sec1C}C: Additional nanofabrication}
Following stack transfer, the heterostructure was shaped into a bar (with additional extended regions above the SWNT to avoid etching it) by reactive ion etching the heterostructure with CHF$_3$ through a resist mask defined by e-beam lithography. A second lithography step defined the graphene contact electrodes [Fig. \ref{fig:S2}(h)], which were made by reactive ion etching to expose a clean graphene edge, and then depositing a metallic trilayer (2 nm Cr/8 nm Pd/50 nm Au) using thermal evaporation (as described in \citeS{S_Wang2013}). A final device is shown in [Fig. \ref{fig:S2}(i)].

We briefly note that it is possible to invert the order of the layers in the SWNT-graphene heterostructure, so that the hBN-encapsulated graphene is placed on the substrate first, and then a CNT is transferred on top. This improves the quality of the graphene by enabling the use of a thicker hBN layer between the graphene and the SiO$_2$. However, the extremely low friction between SWNTs and clean hBN can result in the SWNT bending and shifting from the intended transfer position, sometimes shorting to exposed parts of the graphene or breaking due to stress in fabrication. In contrast, transferring the SWNT to SiO$_2$ pins the SWNT to the rougher substrate, and allows for screening out poorly-conducting SWNTs before transferring a stack. It was thus found to be a more reliable fabrication method, although the yield of working devices after SWNT transfer was still low due to poor electrical contact to the SWNT, SWNTs breaking during stack transfer, and the thin, bottom hBN flake shifting or cracking during stack transfer and allowing the graphene to short the SWNT.

\section{\label{sec2a}II: SWNT conductance}
This section is a brief note on the transport behavior of the SWNTs in our devices. In Figure 1(c), the conductance of the SWNT decreases with increasing gate voltage $V_g$, characteristic of a hole-doped nanotube. There is also a notable dip in the conductance around $V_g=40$ V. While at first this may appear to be the charge neutrality point of the metallic SWNT, it was found to shift its position depending on the direction of the gate voltage sweep, as shown in Figure S3(a). However, the overall SWNT conductance away from the dip generally decreased with increasing $V_g$, regardless of the gate sweep direction. Furthermore, the drag resistance measured during the same gate voltage sweeps does not change polarity when the SWNT conductance dip is on the left side of the graphene CNP as opposed to on the right (Fig. S3(b)). We therefore conclude that the SWNT appears to remain hole-doped at all accessible $V_g$ values. The most likely origin of this conductance dip shift is doping from charges on the substrate, which are reconfigured as the silicon gate is kept at a particular voltage (as before the start of a gate sweep) and then swept. The uncertainty in the nanotube CNP position is incorporated into estimates of the SWNT Fermi energy plotted in the main text (Fig. 3(b) of main text). More details of the calculation are in Section VI below.

Due to the high hole density in the SWNT, the Fermi wavevectors of the SWNT and graphene are significantly mismatched when the drag signal is maximized near the graphene CNP (see Section V for further discussion of drag mechanisms). Although the increased electron-electron scattering phase space when $k_F^{Gr} = k_F^{NT}$ may generically lead to an enhancement $R_{drag}$, we expect that the high $k_F$ where this would occur would mean the Coulomb interaction is already quite suppressed, reducing the drag signal.

In some devices, we do observe a change in the polarity of the drag resistance on either side of a dip in SWNT conductance. We provide an example in a double-gated bilayer graphene (BLG)-SWNT drag device, also used to estimate capacitance ratios in Supplemental Material section VI. Figure \ref{fig:BLG}(a) shows the SWNT conductance tuned by the back gate (the top gate has almost no effect on the SWNT, as it is generally screened by the graphene) and (b) is a color plot of drag resistance as a function of back gate and top gate voltages. There is a clear sign reversal with both the graphene CNP (tuned by both top and back gates and thus diagonal on this plot) and SWNT conductance dip (tuned only by the back gate). This double sign reversal allows us to say with some confidence that the SWNP conductance dip is in fact the CNP.

\section{\label{sec2}III: DC measurement technique}
Low-frequency AC measurements are a standard technique for electrical transport experiments, including many Coulomb drag and other double-layer measurements (e.g. Refs. \cite{S_Liu2017,S_Li2016}). However, drag measurements are sensitive to parasitic effects that can obscure the behavior that truly arises from interlayer charge carrier interactions \cite{S_Gramila1991,S_Hill1996}. For example, when a bias is applied to the drive layer to initiate current flow, the layer acquires a non-zero potential due to contact and layer resistances of the drive layer, which may cause an asymmetric gating effect on the drag layer. This is particularly problematic for AC measurements, since the alternating potential on the drive layer can capacitively couple to the drag layer and generate an alternating current, which can cause a spurious alternating voltage signal in the drag layer due to its contact and layer resistances \cite{S_Hill1996}. 

The interlayer potential can be manually adjusted to approximately zero using resistance bridges \cite{S_Gramila1991,S_Hill1996}. However, as the gate voltage is changed during the measurement, the layer resistances change substantially, unbalancing the bridge circuit. Another problem in our system is the disparity in contact resistances between the graphene and SWNT; while graphene contacts typically have resistances on the order of 100 $\Omega$, the lowest SWNT resistances we achieved in these devices were $\sim 100$ k$\Omega$. This renders the SWNT-graphene drag devices very susceptible to interlayer bias effects.

DC measurements avoid introducing spurious signals from capacitive coupling but are more sensitive to noise and must be carefully monitored to ensure the signal in the drag layer is not purely thermoelectric in origin, as addressed in the main text. To circumvent these issues, we performed our measurements by symmetrically DC biasing the drive layer, applying $\pm V_{drive}/2$ to either side of the SWNT or to a pair of contacts in the graphene (adding a pair of resistors in series to limit the current; the resistor values can be modified, or an additional resistor inserted, to account for differences in contact resistance) and measuring the resulting $I_{drive}$. The circuits are shown in Figure S3. The value of $V_{drive}$ (and thus $I_{drive}$) is swept through a range of values (typically $V_{drive}=0.3 \;\textnormal{V} \rightarrow -0.3\; \textnormal{V}$ on one contact, and $-0.3\; \textnormal{V} \rightarrow 0.3\; \textnormal{V}$ on the other) multiple times, and the detected $V_{drag}$ values at each $I_{drive}$ were averaged to reduce noise. The error for each displayed $V_{drag}$ data point is calculated using the standard error on the mean: $\sigma_{\bar{x}}=\sigma/\sqrt{n}$, where $\bar{x}$ represents the average of a population of measurements, $\sigma$ is the standard deviation, and $n$ is the number of independent measurements. 

\section{\label{sec2b}IV: Additional Onsager data}
We present here additional data comparing the drag response using the two different circuit configurations: driving current in the SWNT and measuring voltage across two probes on graphene, and driving current in the graphene and measuring voltage across the SWNT.

Figure \ref{fig:OnsC} shows a color plot of drag voltage versus drive current and gate voltage at $T=300$ K, using the graphene as the drive layer and SWNT as the drag layer. We have applied a narrow ($\pm3$ V range) moving average filter in the x-direction (gate voltage axis), essentially binning the data from adjacent gate voltages together, to reduce noise and make overall trends in the data more apparent. The same general behavior is apparent in this reciprocal configuration as shown in the main text for SWNT-drive, graphene-drag measurements (Figure 1(e) of the main text), although the higher degree of noise in measurements using the SWNT as the drag layer complicates a straightforward comparison based on the color plots alone.

Additional examples of drag voltage versus drive current data for reciprocal drag circuit configurations at individual gate voltages are shown in Figure \ref{fig:OnsL}. Our moving average filter is restricted to an even smaller range ($\pm1$ V) for better quantitative comparison. The range of $I_{drive}$ is restricted to avoid nonlinearites due to high bias, although some nonlinearity in the SWNT-drive data (orange, filled circles) at gate voltages nearer the graphene CNP (e.g. Fig. \ref{fig:OnsL}(c-e)). Although there is more noise in the measurements for which the SWNT was the drag layer (blue, open circles), the general degree to which the two data sets coincide and the overall linearity of both measurements provide strong evidence that Onsager reciprocity is respected in this regime.

The higher level of noise in the measured drag voltage when the SWNT is the drag layer, which is consistently observed across all gate voltages and temperature, merits some discussion. There are a few potential reasons for increased voltage fluctuations when the SWNT is the drag layer. First of all, the drag resistance is on the order of a few Ohms, while the channel resistance of graphene is less than $\sim$ 1 k$\Omega$, and the resistance of the SWNT is larger than 1 M$\Omega$. Such a large disparity between the SWNT resistance and the drag resistance makes it challenging to detect small resistance variations when we use the SWNT as drag layer. The higher resistance of the SWNT means that small current fluctuations result in larger voltage noise than would be observed if another graphene sheet were used as the drive layer. 

An additional contributing factor may be random reconfiguration of mobile charges on the SiO$_2$ substrate. In our device geometry, the graphene channel is encapsulated, but the SWNT is in direct contact with the SiO$_2$. Thus, stochastic charge fluctuations in the charge traps on the substrate can induce voltage fluctuations in the SWNT when it is being used as drag layer to probe potential. For SWNT drive, we apply a relatively large bias voltage to obtain the same amount of drive current, thus usually fluctuations in the charge environment would not affect the driving current. The graphene drag layer is much less disordered and thus less susceptible to charge fluctuations on the SiO$_2$.

Finally, the increased noise level in the graphene drive-nanotube drag configuration may be a result of the high SWNT resistance relative to the amplifier impedance. The input impedance of the voltage amplifier we use is 100 M$\Omega$ (SR560) while the load resistance of nanotube is $\sim$2 M$\Omega$. In order to avoid spurious capacitive coupling, we measure the DC drag at slow speed, corresponding to 300 ms of averaging time and $\sim$1 second per data point acquisition. In this regime of low frequency measurement with large source resistance, the noise figure (NF) of our preamp is $\sim$0.5 dB (https://www.thinksrs.com/products/sr560.htm), suggesting that the noise is completely dominated by Johnson noise across the source resistance. The effective rms noise is $\sim$0.5 $\mu$V, comparable to the fluctuation of the data we observed in the graphene drive-nanotube drag measurement. Note that for the nanotube drive-graphene drag measurement, the source resistance drops to $\sim$1 k$\Omega$, where the fluctuation in the measured signal is dominated by amplifier noise (NF $\sim$20). We estimate that the rms noise for this measurement is $\sim$20 nV due to the reduced source resistance despite the increased amplifier noise.

\section{\label{sec3}V: Graphene chemical potential and its effect on Coulomb drag}
This section describes a method to qualitatively relate the charge carrier density dependence (more precisely, the gate voltage dependence) of $R_{drag}$ in the SWNT-graphene system by comparison to perturbation theory for graphene-graphene drag \citeS{S_Narozhny2012,S_Peres2011}. Although this theory cannot account for the mixed-dimensional nature of our devices, its applicability in a broad range of carrier densities and temperatures make it a useful framework to understand some of the behavior in our system. Furthermore, the discrepancies between the 2D-2D theory and 1D-2D experiment may illuminate the ways in which dimensionality plays a role in the Coulomb drag behavior.

In comparing our experimental data with theoretical models, it is critical to know the charge carrier densities in the SWNT and graphene ($n_{NT}$ and $n_{Gr}$), as well as their relationship with the chemical potential $\mu$ in each conductor. The net charge carrier density of single-gated graphene heterostructures is generally well approximated by a parallel-plate capacitor formula:
\begin{equation}
    n_{Gr}=\frac{C(V_g-V_0)}{e},
\end{equation}
where $V_0$ is the gate voltage of the graphene charge neutrality point (CNP) and $C$ is the capacitance per unit area between the gate and graphene. However, this formula ignores the effect of the metallic SWNT on the local electrostatic environment of the graphene. Similarly, we cannot simply apply the analytic formula for the capacitance between a wire and a conducting plane \cite{S_Ilani2006} to estimate $n_{NT}$ because the nearby graphene will substantially screen the electric field from the back gate, even though it is not situated between the gate and SWNT. In addition, our experiments are performed at relatively high temperature, so the approximation $\mu \approx E_F$ is not necessarily valid.  

A more accurate approach to modeling the carrier densities and chemical potentials of the two layers (neglecting momentarily the local effect of the SWNT on the nearest region of the graphene) starts with the coupling between the back gate and the conductors, given by:
\begin{subequations}
\label{allequations}
\begin{eqnarray}
eV_g=\mu_{Gr}+\frac{e^2 (n_{Gr}+n_{NT}/2\pi r_{NT})}{C_1} \label{equationa} \\
\mu_{Gr}=\mu_{NT} + \frac{e^2 n_{NT}}{C_2}\label{equationb}
\end{eqnarray}
\end{subequations}
where $C_1$ is the capacitance per unit area between the graphene and back gate, $C_2$ is the capacitance per unit length between the SWNT and graphene, and $r_{NT}$ is the radius of the SWNT. This is analogous to the single-gated graphene-graphene drag device considered in \cite{S_Narozhny2012,S_Peres2011,S_Kim2011}. The SWNT takes the role of the “top” (more heavily screened) layer despite its position between the graphene and back gate because it is much closer to the graphene than to the back gate (2 nm versus 1 $\mu$m). In contrast, the screening effect of the SWNT on the graphene is significant but confined to $< 10$ nm on either side of the SWNT (see Fig. S5 and discussion of a finite element method based on COMSOL simulations in the following section for more details). Figure S4 schematically illustrates the effect of the back gate on the graphene and SWNT bands.

To verify that $n_{Gr}$, $n_{NT}$ are linearly proportional to $V_g$, we can find the density $n_{Gr}^*$ at which the electrical and chemical potentials become comparable:
\begin{equation}
    \hbar v_F \sqrt{\pi n_{Gr}^*}=\frac{e^2 n_{Gr}^*}{C_1}\rightarrow n_{Gr}^* \approx 7\times 10^6\;  \textnormal{cm}^{-2},
\end{equation}
which is several orders of magnitude smaller than the residual impurity density, $\delta n\sim 1.1\times 10^{10}\; \textnormal{cm}^{-2}$, and corresponds to $V_g=0.3$ mV, smaller than the resolution of our data points. Thus, the quantum capacitance of graphene can be neglected in our experiment, and $V_g\propto n_{Gr}$ for all densities under consideration. Furthermore, the SWNT carrier density does not seem to substantially affect the drag behavior, apart from reducing the magnitude of $R_{drag}$ as $n_{NT}$ increases (evident in lower magnitude of the h-h drag on the left side of the graphene CNP compared to the e-h drag on the right side of the graphene CNP, e.g. in main text Figure 1(f)). We observe essentially similar drag behavior regardless of the position of the SWNT conductance dip discussed in Section II, as long as it does not overlap with the CNP of the graphene channel. Since experimentally, the SWNT is observed to be heavily p-doped, and remains degenerate, $V_g\propto n_{NT}\propto \mu_{NT}$, considering the constant density of states of the 1D SWNT band structure \cite{S_Wong2011}. It should also be noted that, due to the chiral nature of graphene, the enhancement of electron-electron scattering at matched Fermi wavevectors (when $q=2k_F$) that is typical in 2D electronic systems is not present in graphene-graphene drag systems \cite{S_Hwang2011}. Investigation of such enhancement in the SWNT-graphene system requires low-doping nanotubes and transport measurements near the nanotube CNP, an avenue for future study.

We can now consider 3 possible regimes of drag response, based on $\mu_{Gr}$ and $T$. When both are small ($\mu_{Gr}$, $k_B T$ $< \tau^{-1}$, where $\tau(\mu_{Gr},T)$ is the scattering time), the transport is disorder-dominated and $n_{Gr}$ becomes temperature-independent \cite{S_Narozhny2012}:
\begin{equation}
    n_{Gr}(\mu_{Gr},k_B T < \tau^{-1})=\frac{\mu_{Gr}}{\hbar v_F^2 \tau}.
\end{equation}

The relevant expression for Coulomb drag in this regime is either
\begin{equation}\label{drag2l}
    \rho_{drag}(\mu_{Gr},\mu_{NT}\ll T) \approx 1.41 \alpha^2 \frac{\hbar}{e^2} \frac{\mu_{Gr} \mu_{NT}}{k_B^2 T^2},
\end{equation}
where $\alpha=e^2/v_F$ is the interaction strength, or at higher $\mu_{NT}$ (with $\mu_{NT}>\mu_{Gr}$ in our experiment),
\begin{equation}
    \rho_{drag}(\mu_{Gr}\ll k_B T \ll \mu_{NT}) = 5.8\alpha^2 \frac{\hbar}{e^2} \frac{\mu_{Gr}}{\mu_{NT}}.
\end{equation}

In either case, $\rho_{drag}\propto \mu_{Gr}$, and since we have established that $\mu_{Gr}\propto n_{Gr} \propto V_g$ in this regime, the theory predicts $\rho_{drag}\propto V_g$. This prediction agrees with our $R_{drag}$ data for gate voltages close to the graphene CNP [see for example Fig. 1(d),(f) of the main text]. Due to the high SWNT hole density in the accessible gate voltage range in our devices, as discussed in Section II, we expect that the regime of Equation \ref{drag2l} is not observed in our experiments.

At higher $T$, the graphene charge carriers close to the CNP ($T\tau>1$, $\mu_{Gr}\ll T$) are in the Dirac fluid regime. In this case, the carrier density remains linear in $\mu$ but acquires temperature dependence:
\begin{equation}
    n_{Gr}(\mu_{Gr}\ll k_B T)=\frac{1}{\hbar^2 \pi v_F^2} 4 \textnormal{ln}(2) \mu_{Gr} k_B T.
\end{equation}
A similar relationship between $\rho_{drag}$ and $V_g$ holds as in the disorder-dominated regime, although the temperature dependence is reduced by a factor of $T$ (since now $\mu_{Gr}\propto n_{Gr}/T$). However, since we have relatively few data points in this regime, it is difficult to conclusively compare the theoretical and experimental temperature dependences.

Finally, $\mu_{Gr}\gg k_B T$ is the Fermi liquid regime, where $n_{Gr}$ once again loses its temperature dependence:
\begin{equation}
    n_{Gr}(\mu_{Gr} \gg k_B T) = \frac{1}{\hbar \pi v_F^2} \mu_{Gr}^2.
\end{equation}
We must also consider the Fermi liquid expression for the Coulomb drag:
\begin{equation}
    \rho_{drag}(\mu_{NT}>\mu_{Gr} \gg k_B T) \approx \alpha^2 \frac{\hbar}{e^2} \frac{8\pi^2}{3} \frac{k_B^2 T^2}{\mu_{Gr}\mu_{NT}} \textnormal{ln}\left( \frac{\mu_{Gr}}{k_B T}\right),
\end{equation}
which implies $\rho_{drag}\propto T^2/V_g^{1/2}$, neglecting any contribution from the SWNT. While this is qualitatively similar to the experimental behavior of $R_{drag}$ at higher $V_g$ (i.e. increasing with temperature and decaying as a power law with $V_g$), it does not align with a more quantitative analysis of the data (which finds $R_{drag} \propto V_g^{-1}$ to $V_g^{-1.3}$ at $140 \; \textnormal{K} < T < 300 \; \textnormal{K}$; see Fig. 3(c) of the main text). This disagreement suggests a more detailed theoretical analysis of the SWNT-graphene drag system is required to fully understand the mechanisms at play.

\section{\label{sec4}VI: Estimation of SWNT and graphene carrier densities}
This section expands on the estimations of SWNT and graphene carrier densities discussed in the first part of the previous section. As stated, straightforward application of the analytical formulae for capacitance between parallel conducting planes (for graphene-back gate capacitance) or a wire and a ground plane (for SWNT-back gate capacitance) would not adequately account for the electrostatic environment of either the SWNT or graphene. As such, we used COMSOL Multiphysics to perform a finite-element analysis of the gate, conductors, and dielectrics and computed the induced charge on the SWNT and graphene due to an applied gate voltage. Material parameters such as the relative permittivity were found in Refs. \cite{S_Dean2010,S_Wong2011,S_Ng1995,S_Zhang2017,S_Laturia2018,S_Seol2010,S_Elias2011,S_Nicklow1972,S_Lu2007,S_Kozinsky2006}. Figure \ref{fig:S5} shows a color map of the electric field in this region due to an applied $V_g=1$ V, along with the resulting charge carrier density in the graphene. The simulation suggests that the local electric field (and thus the local carrier density) is reduced by a factor of $\sim$30 in the graphene directly above the SWNT compared to the value predicted from a parallel-plate capacitor model. Due to the local nature of the Coulomb interaction ($\propto r^{-3}$), we expect that this region of decreased carrier density is the part of the graphene that most directly contributes to Coulomb drag. We use the capacitance of this screened region as a lower bound in Figure 3(b) of the main text, while retaining the analytic “bulk” value as an upper bound [see Fig. \ref{fig:S6} for equivalent capacitance circuits for both scenarios]. Since the area of carrier depletion in the graphene is extremely narrow (the width of this “screening well” is $\sim$7 nm, beyond which the graphene carrier density rapidly approaches the value predicted by the analytical model), contributions to drag from higher-$n_{Gr}$ regions could also be important.

The same simulation was used to estimate the capacitance (and thus carrier density) of the metallic SWNT; this is its lower bound in Figure 3(b) in the main text. Since this value may be sensitive to material parameters, such as the dielectric permittivity of the SWNT, the precise magnitudes of which we do not know, we need an alternative approach to validate our calculation and estimate the possible range of the capacitance value. For this purpose, we employ gate dependent data from SWNT/bilayer graphene (BLG) [Fig. \ref{fig:S7}(a)]. Here the SWNT acts as a local gate on BLG, which was also coupled to a gold top gate and silicon back gate. Since the device geometry is close to the SWNT/monolayer graphene devices in which we measure drag, electrostatic measurements from the SWNT/BLG device allow us to infer the capacitive coupling of the SWNT to monolayer graphene in the drag devices. Particularly, by comparing the effects of the top and back gates on the BLG with the effect of the SWNT gate, we can determine the degree of SWNT-BLG coupling, and therefore how to account for the proximity of the BLG. We then rescale the results to account for the very slightly different geometry of the devices being considered in the main text of this paper.

For the SWNT-BLG device shown, which has a 4 nm-thick hBN flake separating the SWNT and BLG and 1 $\mu$m of SiO$_2$ between the heterostructure and back gate, the BLG resistance as a function of back and top gate voltages ($V_{bg}$ and $V_{tg}$) is shown in Figure \ref{fig:S7}(b). The slope of the line tracking the BLG charge neutrality point as the gate voltages are changed, $dV_{bg}/dV_{tg}=-14.66$, quantifies the strength of the capacitive coupling to the top gate compared to the back gate ($dV_{bg}/dV_{tg}=C_{tg}/C_{bg}$). Similarly, tracking the position of the side-peak caused by local SWNT gating while also varying the top gate [Fig. \ref{fig:S7}(c)] gives the relative coupling of the BLG to the SWNT and top gate, $dV_{tg}/dV_{NT} =C_{NT}/C_{tg} =-4.75$.

We can model the effective geometric capacitance of the SWNT to ground as a series combination of back gate and BLG coupling [Fig. \ref{fig:S6}(a-b)]:
\begin{equation}
    C_\textnormal{eff}^{-1} = C_{NT-bg}^{-1}+C_{NT-BLG}^{-1}.
\end{equation}
The first term can be calculated using the formula for wire-plane capacitance per unit length:
\begin{equation}
    \frac{C_{NT-bg}}{L} = \frac{2\pi \epsilon_0 \epsilon_r}{\textnormal{arccosh}\left( \frac{2d_{bg}}{d_{NT}} \right)},
\end{equation}
where $d_{bg} = \;1 \mu$m is the SiO$_2$ thickness, $d_{NT}$ = 2 nm is the SWNT diameter, and $L$ is the length of the SWNT. The second term is estimated by the experimentally-determined coupling of the BLG to the gate and SWNT:
\begin{equation}
    \frac{C_{NT-BLG}}{A}=\frac{C_{bg-BLG}}{A}\frac{dV_{bg}}{dV_{NT}} = \frac{C_{bg-BLG}}{A}\frac{dV_{bg}}{dV_{tg}}\frac{dV_{tg}}{dV_{NT}},
\end{equation}
where $C_{bg-BLG}/A= \epsilon_0 \epsilon_r/d_{bg}$ is the standard parallel-plate capacitor formula.

At this point, we must account for the difference in hBN thicknesses between the BLG device (4 nm) and the SWNT-monolayer graphene drag device (2 nm). Noting that the dielectric thickness enters the wire-place capacitance formula as $1/\textnormal{arccosh}(2d_{BN}/d_{NT})$, reducing the hBN thickness from 4 to 2 nm simply requires multiplying the BLG device result by $\textnormal{arccosh}(2\times 4\textnormal{nm}/2\textnormal{nm})/\textnormal{arccosh}(2\times 2\textnormal{nm}/2\textnormal{nm})\approx 1.567$. Multiplying our adjusted $C_{NT-BLG}/A$ by $d_{NT}$ to convert it to capacitance per unit SWNT length, we find $C_\textnormal{eff}/L=6.11\times10^{-12}$ F/m for the geometric capacitance.

Finally, we must also account for the contribution from the quantum capacitance, which is particularly important for the SWNT. For a metallic SWNT, this has the simple form \cite{S_Kozinsky2006}
\begin{equation}
    C_Q^{NT}=\frac{8e^2}{hv_F}\approx 310\times 10^{-12} \textnormal{F/m}.
\end{equation}
This is substantially larger than geometrical capacitance estimated above but can easily be included in series with the geometric capacitance to give the total $C_\textnormal{bg-NT}/L=6.00\times10^{-12}$ F/m.

\section{\label{secT}VII: Derivation of nonlinear contributions to the drag voltage}

The close connection between breakdown of Onsager reciprocity and appearance of nonlinear drag response suggests that these behaviors may share the same origin. We first note that driving current $I_{NT}$ in the SWNT can produce a nonuniform temperature profile due to the thermoelectric Peltier effect ($\Delta T_P \propto I_{NT}$) and Joule heating ($\Delta T_J \propto I_{NT}^2$). As the graphene and SWNT are in close proximity, this temperature profile can be transferred from the SWNT to the graphene above, particularly when the gate voltage is tuned near the graphene CNP \cite{S_Song2012,S_Song2013,S_Song2013a}. This energy transfer gives rise to a temperature gradient across the voltage probes in the graphene layer, $\Delta T(x)=\Delta T_P(x) + \Delta T_J(x)$, where $x$ is the distance from SWNT to the voltage probes (fixed for any given pair of probes, and thus omitted in the following analysis).

There are two mechanisms by which $\Delta T$ can produce a nonlinear drag response: (i) generation of a thermoelectric voltage $V_{TE}(\Delta T) = -S \Delta T$, where $S$ is the Seebeck coefficient of graphene and $\Delta T \propto I_{NT}^2$; and (ii) temperature dependent change in the drag resistance $R_{drag}$ producing a nonlinear drag voltage: $V_{NL}(\Delta T) = \frac{1}{2}\frac{dR_{drag}}{dT} \Delta T I_{drive}$. The thermoelectric voltage $V_{TE}(\Delta T) = -S \Delta T$ is a well-known phenomenon in systems with a thermal gradient \cite{S_Blundell2009}. The second nonlinear contribution to the drag voltage, $V_{NL}(\Delta T) = \frac{1}{2}\frac{dR_{drag}}{dT} \Delta T I_{drive}$, merits further discussion.

This expression can be obtained by assuming a small temperature gradient between two voltage probes separated by a distance $L$. The local drag resistivity $\rho_{drag}$ relates the drag layer local electric field $\varepsilon$ and drive current density $j_{drive}$ by $\varepsilon=\rho_{drag} j_{drive}$. Now we consider a small, constant temperature gradient $dT/dx$ between the voltage probes where $x=0, L$ correspond to the respective electrode positions. The temperature difference between the electrodes is then $\Delta T=L dT/dx$. The drag voltage between them can then be obtained (with $T_0$ the temperature of the thermal bath):
\begin{eqnarray}
    V_{drag}=\int_{0}^{L} j_{drive}\rho_{drag}(T(x)) dx = j_{drive}\int_{0}^{L} \rho_{drag}(T(x)) dx \\
     = j_{drive}L\rho_{drag}(T_0)+\frac{1}{2}j_{drive}\frac{d\rho_{drag}}{dT}\frac{dT}{dx}L^2\\
     =R_{drag}I_{drive}+\frac{1}{2}\frac{dR_{drag}}{dT}\Delta T I_{drive} 
\end{eqnarray}

The first term is the typical drag resistance $R_{drag}$ between the voltage probes at $x=0$ and $x=L$, and the second term is the nonlinear contribution (since $Delta T\propto I_{drive}$ or $I_{drive}^2$, as discussed in the main text) due to the temperature dependence of $R_{drag}$.

Allowing heating by both effects mentioned above, we identify terms contributing to $V_{drag}$ that are proportional to $I_{NT}$ ($R_{drag}$ and $V_{TE}(\Delta T_P)$), $I_{NT}^2$ ($V_{TE}(\Delta T_J)$) and $V_{NL}(\Delta T_P)$), and $I_{NT}^3$ ($V_{NL}(\Delta T_J)$). The $I_{NT}$-linear thermoelectric response is energy drag, which is observed to be large in graphene systems when both layers are tuned very near the CNP \cite{S_Song2012,S_Song2013,S_Song2013a,S_Gorbachev2012}, but is otherwise negligible. The presence of both quadratic and cubic nonlinearity in our experimental data, with peaks developed in both $|\gamma|$ and $|\eta|$ at the graphene CNP (see main text Fig. 2(b) and (c)), suggests that, at minimum, the temperature dependence of $R_{drag}$ (i.e. (ii) above) must play a significant role. These effects are significant near the CNP, where $|\frac{dR_{drag}}{dT}|$ exhibits large fluctuations at low temperatures due to disorder (see main text Fig. 2(e)).  The nonlinear contribution becomes more appreciable as the linear drag signal diminishes at lower temperature, and in SWNT devices with larger resistance (including contact resistance; see following section for additional data), which also supports the local heating-induced energy transfer picture.

\section{\label{sec5}VIII: Comparison of individual layer transport and drag resistance in different measurements}
The data presented in the main text were primarily gathered from two separate thermal cycles of the same device (D1), with the first set of measurements (D1-A) occurring shortly after the completion of nanofabrication, and the second set of measurements (D1-B) starting approximately 9 months later. In the main text, the data in Figure 1, Figure 2(a-c), and Figure 5 are from D1-A, and the data in Figure 2(d-e), Figure 3 and Figure 4 are from D1-B. Comparing similar measurements for the two different data sets (for example the drag resistance versus back gate voltage in Fig. 1(f) for D1-A and Fig. 3(d) for D1-B), it is apparent that they qualitatively follow the same behavior, but with some quantitative discrepancies. In particular, the back gate voltage range with an appreciable drag signal is much larger for D1-A ($V_g\sim$ 20 V) than for D1-B ($V_g\sim$ 3 V). This can be attributed to a comparable change in the disorder in the graphene, observed as a change in the CNP position and peak width. Figure S8 shows a direct comparison of the SWNT conductance, graphene resistance, and drag resistance as a function of gate voltage for D1-A [Fig. \ref{fig:S8}(a-c)] and D1-B [Fig. \ref{fig:S8}(d-f)]. In both data sets, the drag signal width directly corresponds to the width of the graphene CNP peak. Since all the preceding discussion about possible physical mechanisms has relied on the interplay of various regimes (including a disorder-dominated regime) rather than specific numerical predictions (e.g. that the peaks occur at a specific Fermi wavevector in either graphene or SWNT), our arguments should apply equally well in both data sets.

As an additional comparison, Figures S9 and S10 show drag resistance data from several other devices. There are a few key differences in the geometry of the various devices. For device pair D2, the graphene was etched into 2 bar segments of differing widths [Fig. \ref{fig:S9}(a)]. D2-1 is 1.9 $\mu$m wide, while D2-2 is 600 nm wide. D1 and D3 have a single bar each [Fig. \ref{fig:S10}(a)]; D1 is 1 $\mu$m wide by 7 $\mu$m long, while D3 is 1.2 $\mu$m wide by 9.7 $\mu$m long. The hBN separating the SWNT from the graphene is significantly thicker for D2 (5 nm versus 2 nm for D1 and 3 nm for D3). Finally, the metal electrodes in D2 contact narrow, protruding sections of graphene (“noninvasive” contacts), while in D1 and D3 they directly contact the bar (“invasive” contacts). We also note that the SWNTs are all metallic, but the chiralities and corresponding diameters are different for each device. The SWNT incorporated into D1 has chiral indices (16,13) and diameter 1.97 nm, the SWNT in D2 has chiral indices (21,5) and diameter 2.26 nm, and the SWNT in D3 has chiral indices (21,15) and diameter 2.45 nm.

Measurements of the drag resistance as a function of the gate voltage for D2-1 and D2-2 show consistently smaller signal than D1, which is reasonable given the larger interlayer separation. The exception is when graphene is used as the drive layer, in which case D2-1 shows a comparatively large signal [Fig. \ref{fig:S9}(b-c)]. Drive/drag layer reciprocity is not observed in the wider bar D2-1, and while it is respected to a degree in the narrower bar D2-2 [Fig. \ref{fig:S9}(e-f)], it breaks down at a higher temperature than in device D1 ($T \sim 200$ K in D2-2, compared to $T \sim 140$ K in D1). These measurements were carried out using a small drive current (200 nA) in an attempt to remain in the linear response regime, and the reported $R_{drag}$ in Figure \ref{fig:S9}(b-c),(e-f) is from a linear fit of the drag voltage versus drive current in this small range. Subsequent measurements with larger drive current [Fig. \ref{fig:S9}(d)] show a mostly quadratic drag current-voltage relationship. It is therefore likely that the breakdown of layer reciprocity is due to an earlier and stronger onset of nonlinear transport effects (discussed in main text), and even measurements with a small drive current may have a substantial nonlinear transport contribution. Furthermore, we note that narrow, noninvasive contacts have been predicted not to thermalize efficiently with the electron system in graphene-graphene drag devices near charge neutrality, leading to a breakdown of layer reciprocity even in the linear response regime \cite{Song2013a}. This detail of the device geometry may also contribute to the behavior seen in the D2 devices. Since the width of the narrower device D2-2 is comparable to the width of the noninvasive graphene contacts, the bar and contact can thermalize more effectively, which allows some degree of layer reciprocity to be preserved.

The geometry of device D3 is similar to D1, and it displays layer reciprocity at comparable temperatures [Fig. \ref{fig:S10}(b)]. The drag resistance qualitatively resembles D1, although with a smaller magnitude. This may be attributed to the increased layer separation in device D3. The graphene quality is similar to D1 [Fig. \ref{fig:S10}(c) versus main text Fig. 1(d)], but the SWNT has substantially higher resistance and appears quite disordered [Fig. \ref{fig:S10}(d)]. The drag signal on the positive side of graphene CNP lacks the distinctive peak of the D1 data, likely because SWNT carrier density and current were lower than the corresponding part of the signal in D1. The high-resistance SWNT, as well as additional inhomogeneity appearing during and after thermal cycles, prevented an extensive characterization of device D3. Nonetheless, the initial data we were able to gather support the explanation of the drag behavior in the main text.

\begin{figure}
    \centering
    \includegraphics[width=0.9\textwidth]{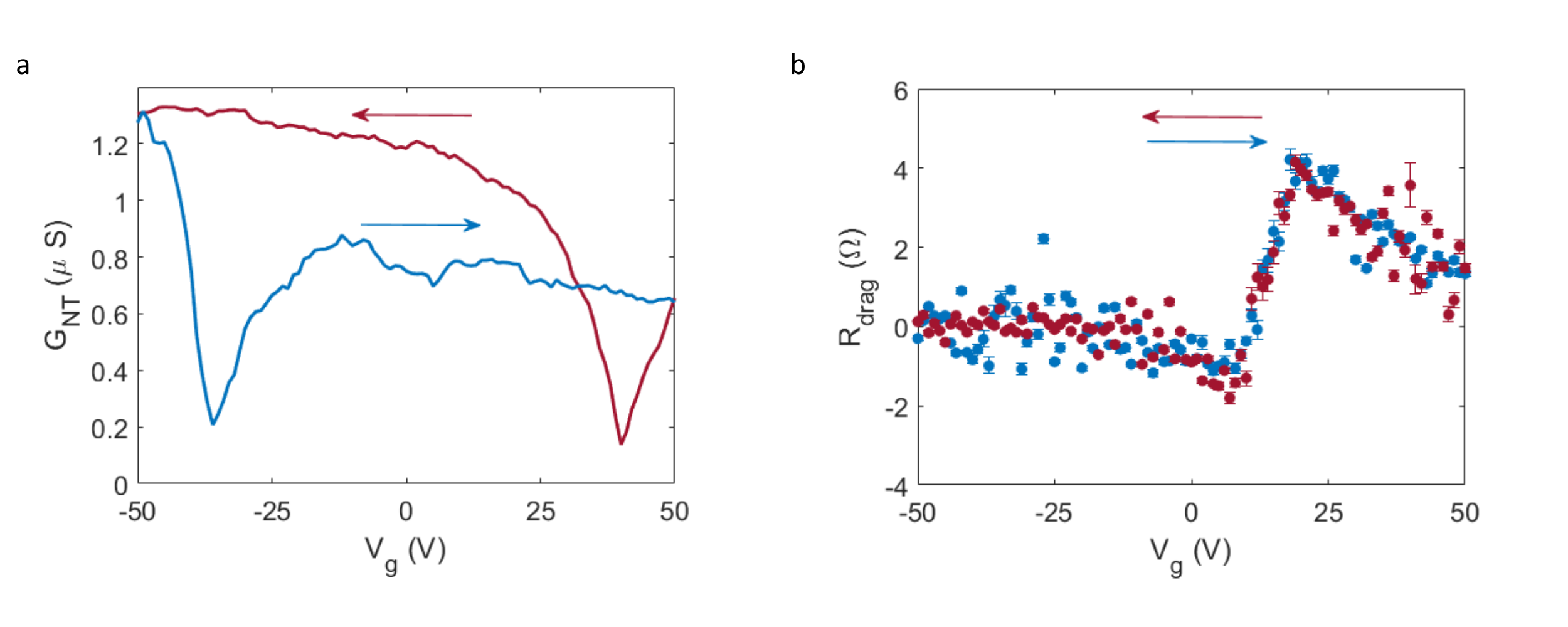}
    \caption{(a) Conductance of the SWNT in device D1 at $T=200$ K as a function of gate voltage. Arrows indicate direction of $V_g$ sweep; positive to negative (red) and negative to positive (blue). (b) $R_{drag}$ measured simultaneously with the SWNT conductance in (a).}
    \label{fig:S2a}
\end{figure}

\begin{figure}
    \centering
    \includegraphics[width=0.9\textwidth]{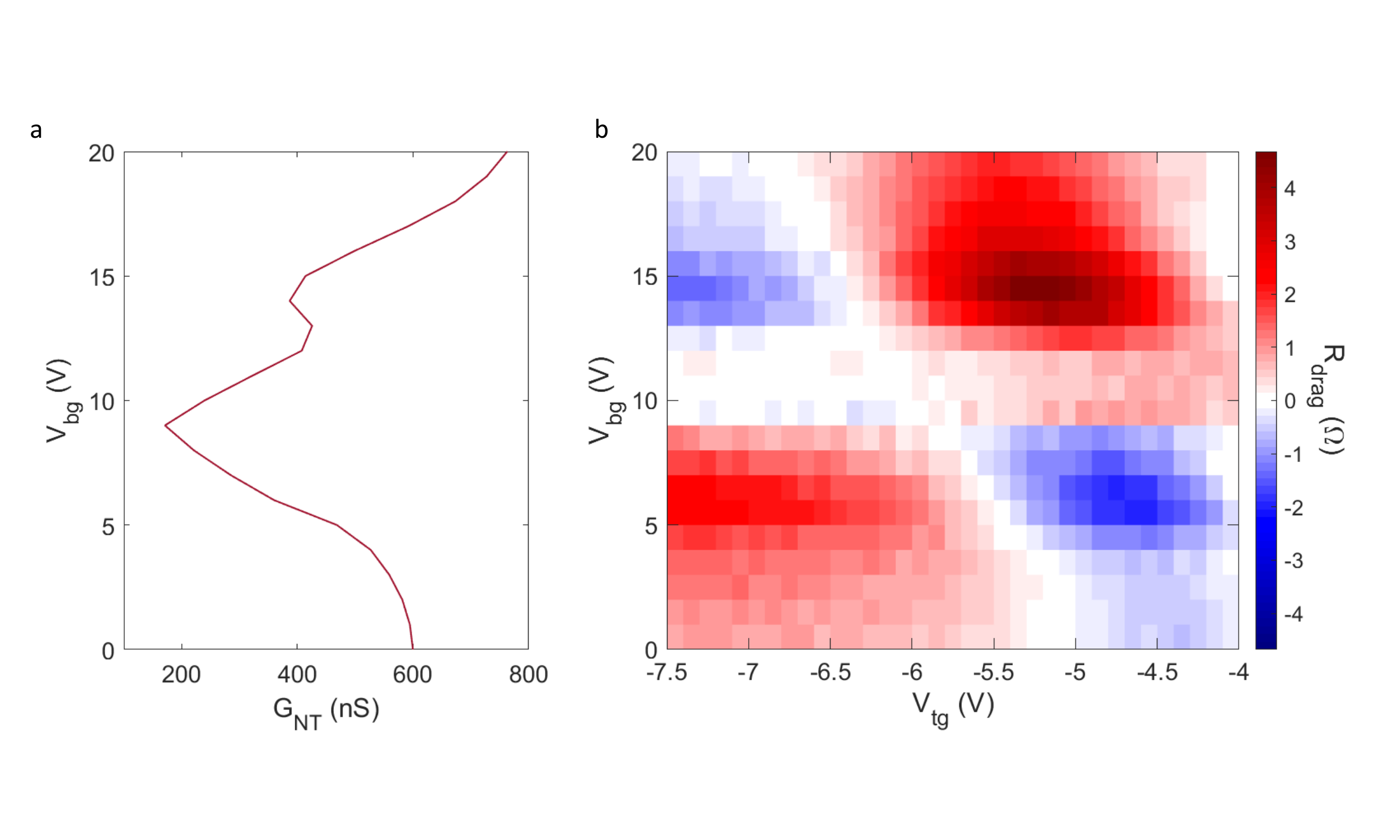}
    \caption{(a) SWNT conductance as a function of back gate voltage in a BLG-SWNT device (axes flipped to align with (b)). (b) Drag resistance as a function of back gate and top gate voltages.}
    \label{fig:BLG}
\end{figure}

\begin{figure}
    \centering
    \includegraphics[width=0.9\textwidth]{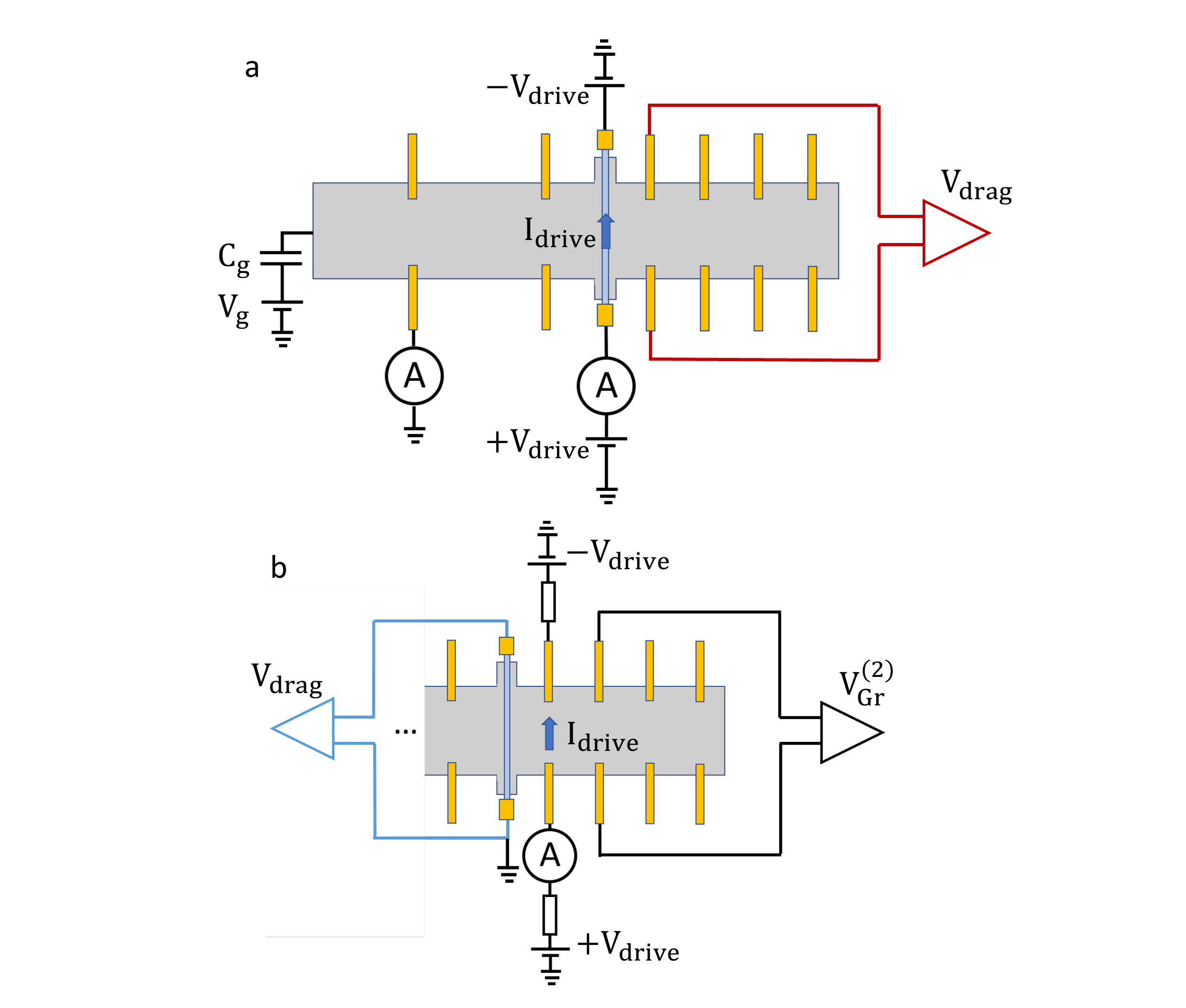}
    \caption{Circuit diagrams for DC measurement. (a) Schematic of symmetrically biased SWNT (drive layer) and voltage measurement on graphene (drag layer). Additional ammeters on SWNT and graphene can measure the SWNT conductance and SWNT-graphene tunneling current, respectively. (b) Symmetrically biased graphene (drive layer) and voltage measurement on graphene (to estimate graphene resistance) and SWNT (drag layer). In this case, large ($\sim10$ M$\Omega$) resistors are added after the voltage sources so that the graphene is being current biased.}
    \label{fig:S3}
\end{figure}

\begin{figure}
    \centering
    \includegraphics[width=0.9\textwidth]{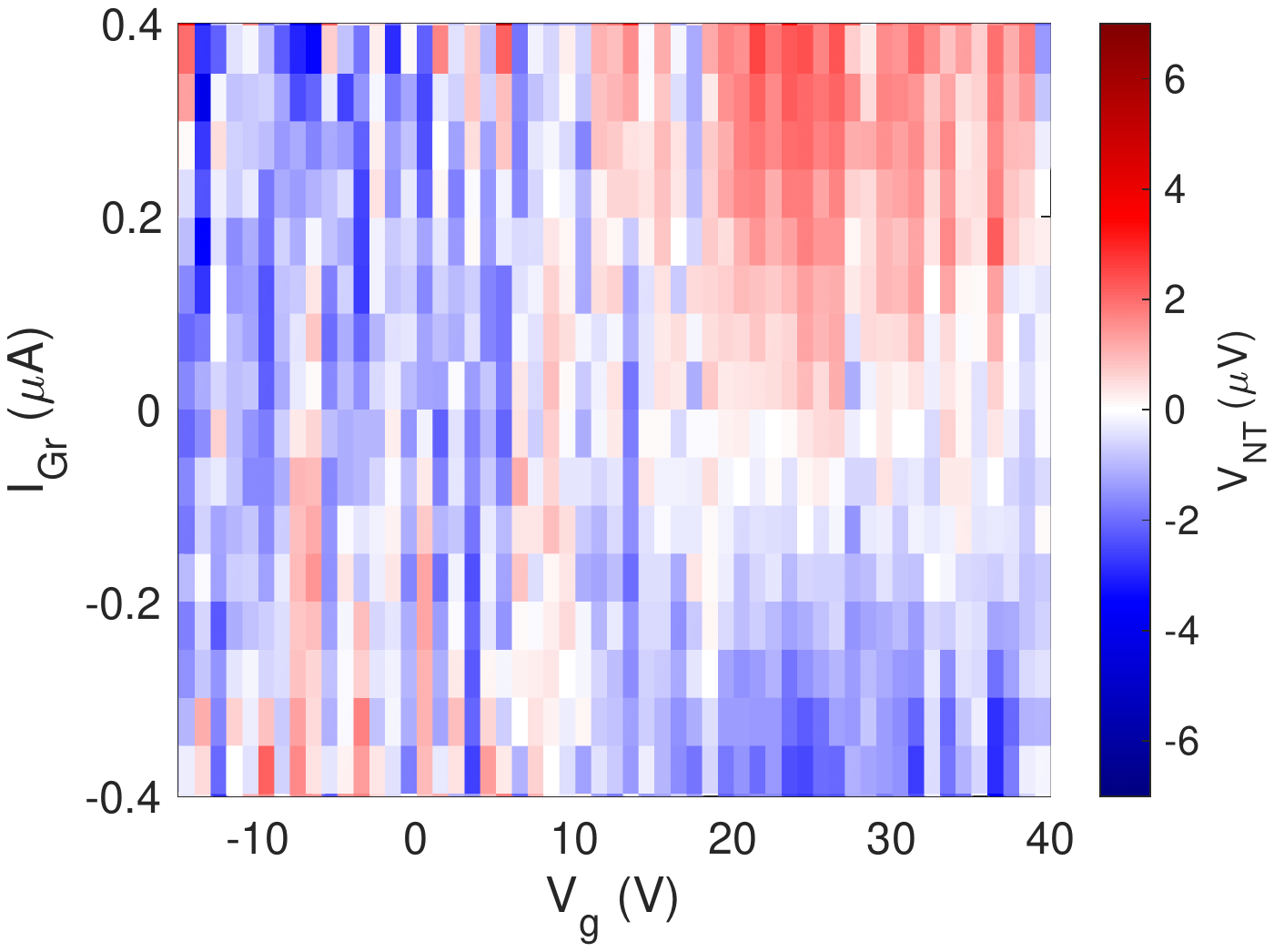}
    \caption{Drag voltage at $T=300$ K as a function of drive current and gate voltage, with graphene as the drive layer and SWNT as the drag layer. Current is applied at the closest pair of voltage probes ($x=800$ nm from the SWNT). Data has been slightly smoothed in the x-direction to reduce noise.}
    \label{fig:OnsC}
\end{figure}

\begin{figure}
    \centering
    \includegraphics[width=0.9\textwidth]{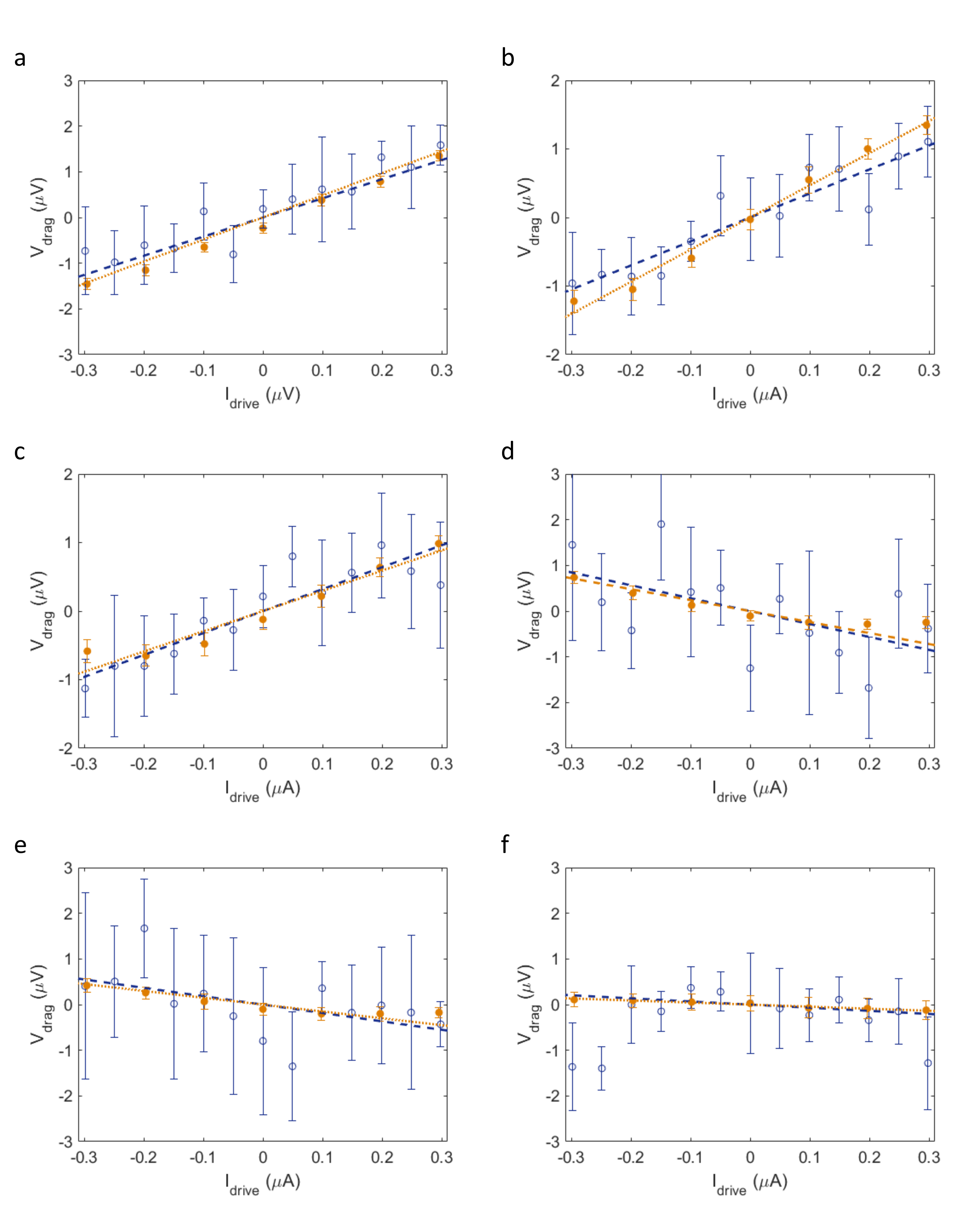}
    \caption{Drag voltage versus drive current at various gate voltages for reciprocal layer configurations: driving current in the SWNT while measuring graphene voltage (orange, filled symbols) and driving current in the graphene while measuring voltage across the SWNT (blue, open symbols) at T = 300 K. $V_g$ value stated is the center of the moving average filter. (a) $V_g = 23$ V. (b) $V_g = 20$ V. (c) $V_g = 16$ V. (d) $V_g = -5$ V. (e) $V_g = -11$ V. (f) $V_g = -18$ V.}
    \label{fig:OnsL}
\end{figure}

\begin{figure}
    \centering
    \includegraphics[width=0.9\textwidth]{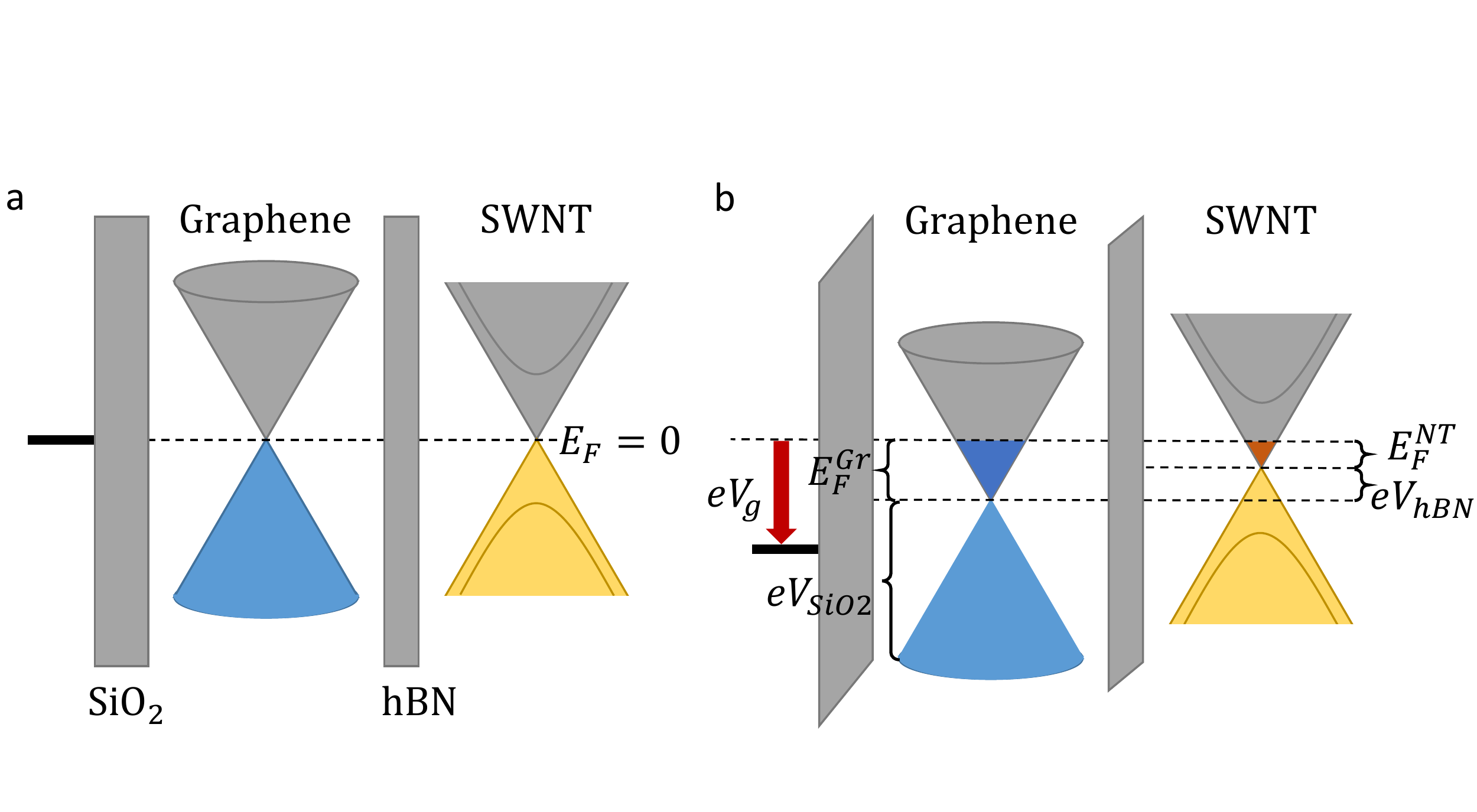}
    \caption{Band diagram of SWNT-graphene heterostructure (a) at $V_g = 0\;\textnormal{V}$ (b) at positive gate voltage. Graphene band filling is represented in dark (light) blue and SWNT band filling is represented in orange (yellow) for conduction (valence) band. Adapted from \cite{S_Kim2011}.}
    \label{fig:S4}
\end{figure}

\begin{figure}
    \centering
    \includegraphics[width=0.9\textwidth]{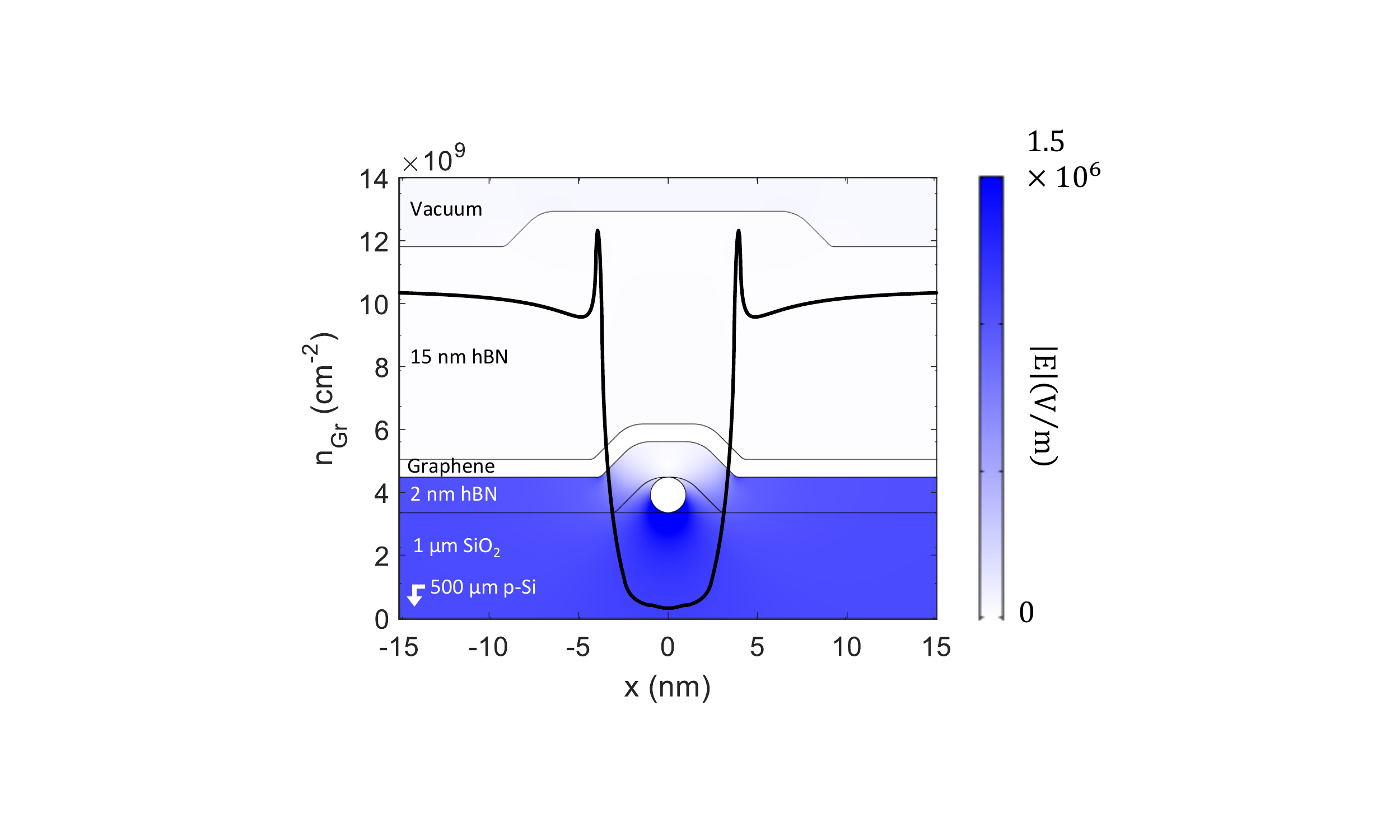}
    \caption{COMSOL simulation of SWNT-graphene device. Color map shows $|E|$ with 1 V applied to back gate for a device cross-section. Overlay (black line) shows calculated graphene carrier density $n_{Gr}$ as a function of distance $x$ from the SWNT.}
    \label{fig:S5}
\end{figure}

\begin{figure}
    \centering
    \includegraphics[width=0.9\textwidth]{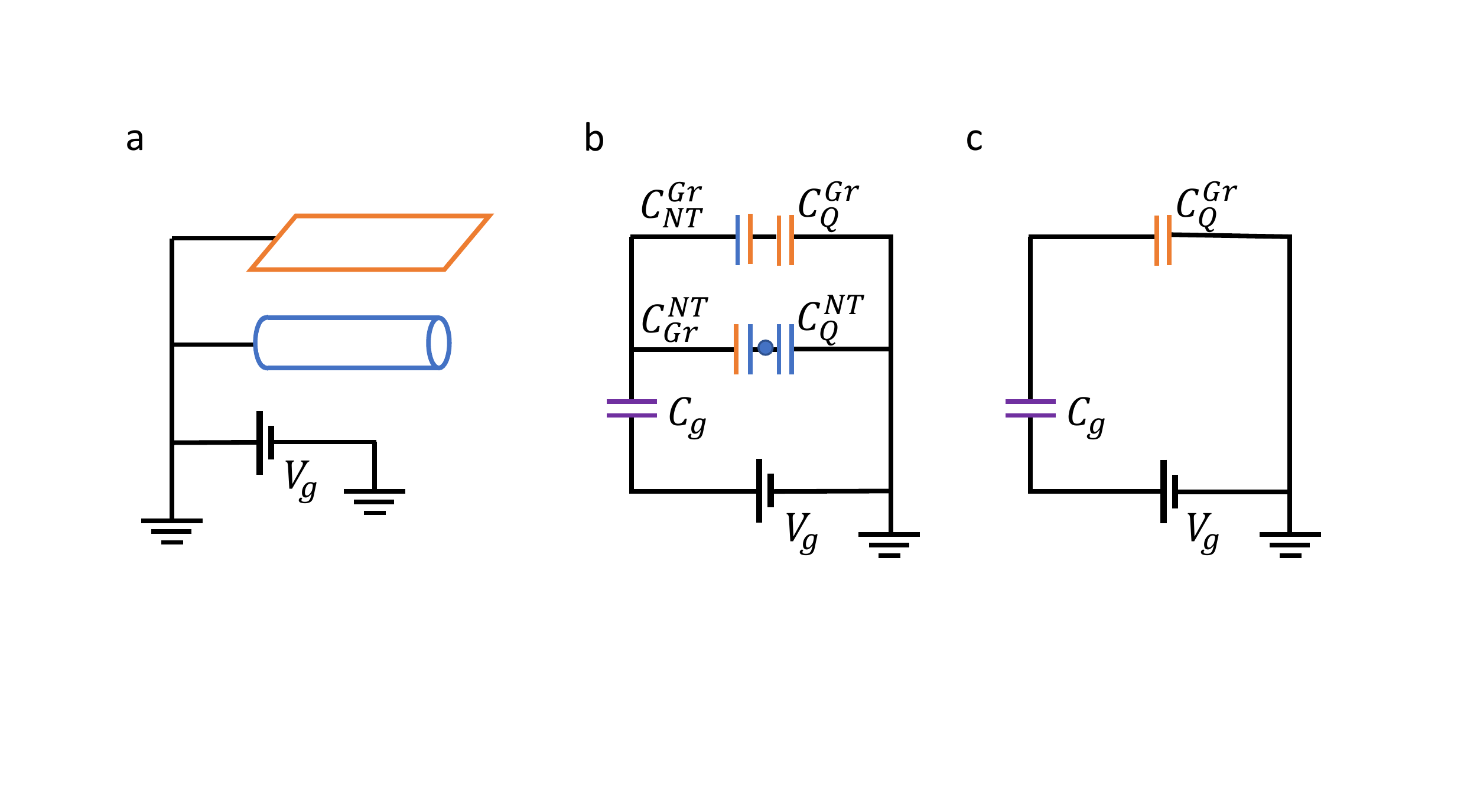}
    \caption{Equivalent capacitance circuit of SWNT-graphene device. (a) Schematic of conductors and gate voltage. (b) Circuit for local screening between graphene and SWNT. SWNT-graphene capacitance is illustrated as two capacitors because each conductor screens some of the gate electric field from the other. (c) Circuit for graphene capacitance far from SWNT.}
    \label{fig:S6}
\end{figure}

\begin{figure}
    \centering
    \includegraphics[width=0.9\textwidth]{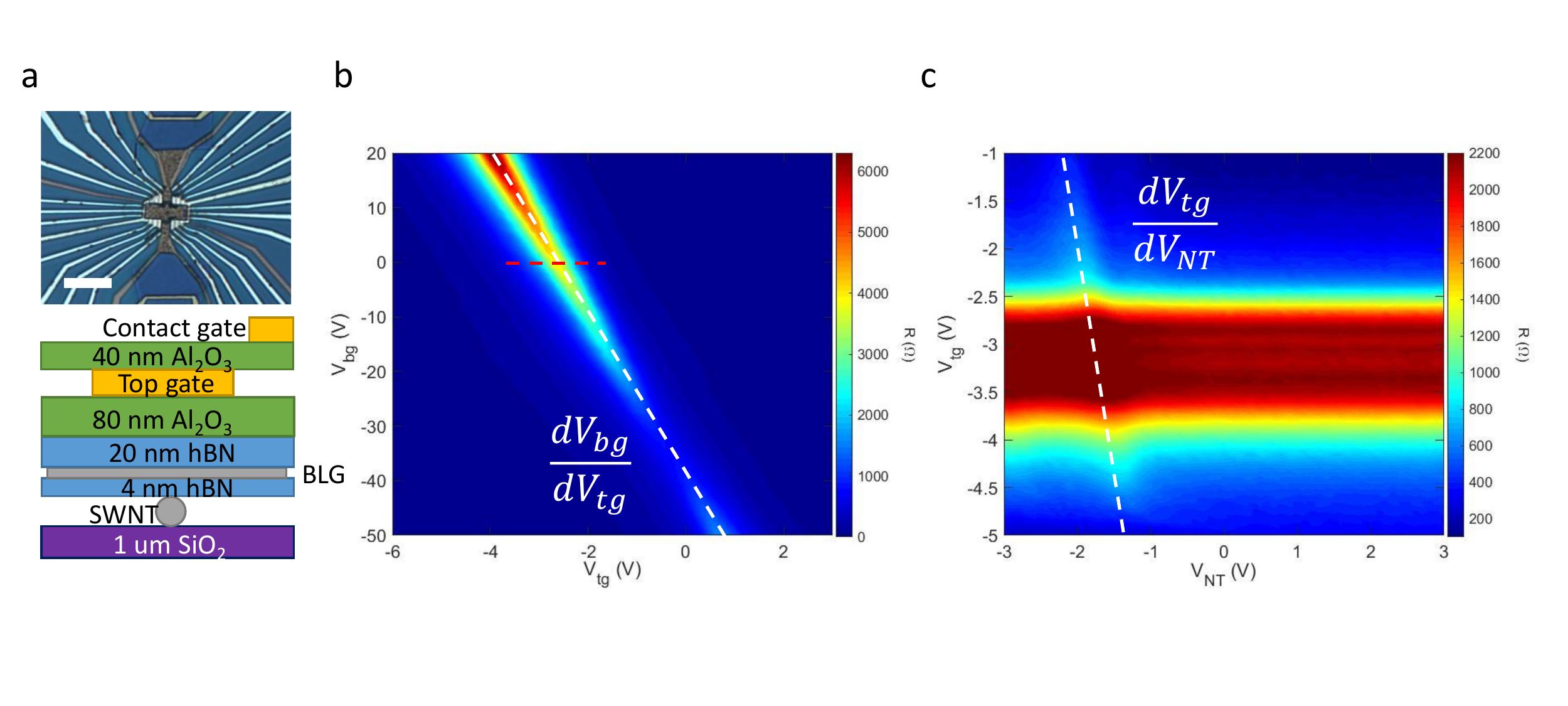}
    \caption{(a) Optical microscopy image (upper panel) and cross-section schematic (lower panel) of SWNT-BLG device. Scale bar is 10 nm. (b) BLG resistance as a function of back gate and top gate voltages. Dashed white line follows the charge neutrality point and has slope $dV_{bg}/dV_{tg}$. Dashed red line indicate back gate voltage (0 V) and top gate voltage range for panel (c). (c) BLG resistance as a function of top gate voltage and SWNT local gate voltage. Dashed white line indicates secondary Dirac peak due to local gating of the BLG by the SWNT and has slope $dV_{tg}/dV_{NT}$.}
    \label{fig:S7}
\end{figure}

\begin{figure}
    \centering
    \includegraphics[width=0.9\textwidth]{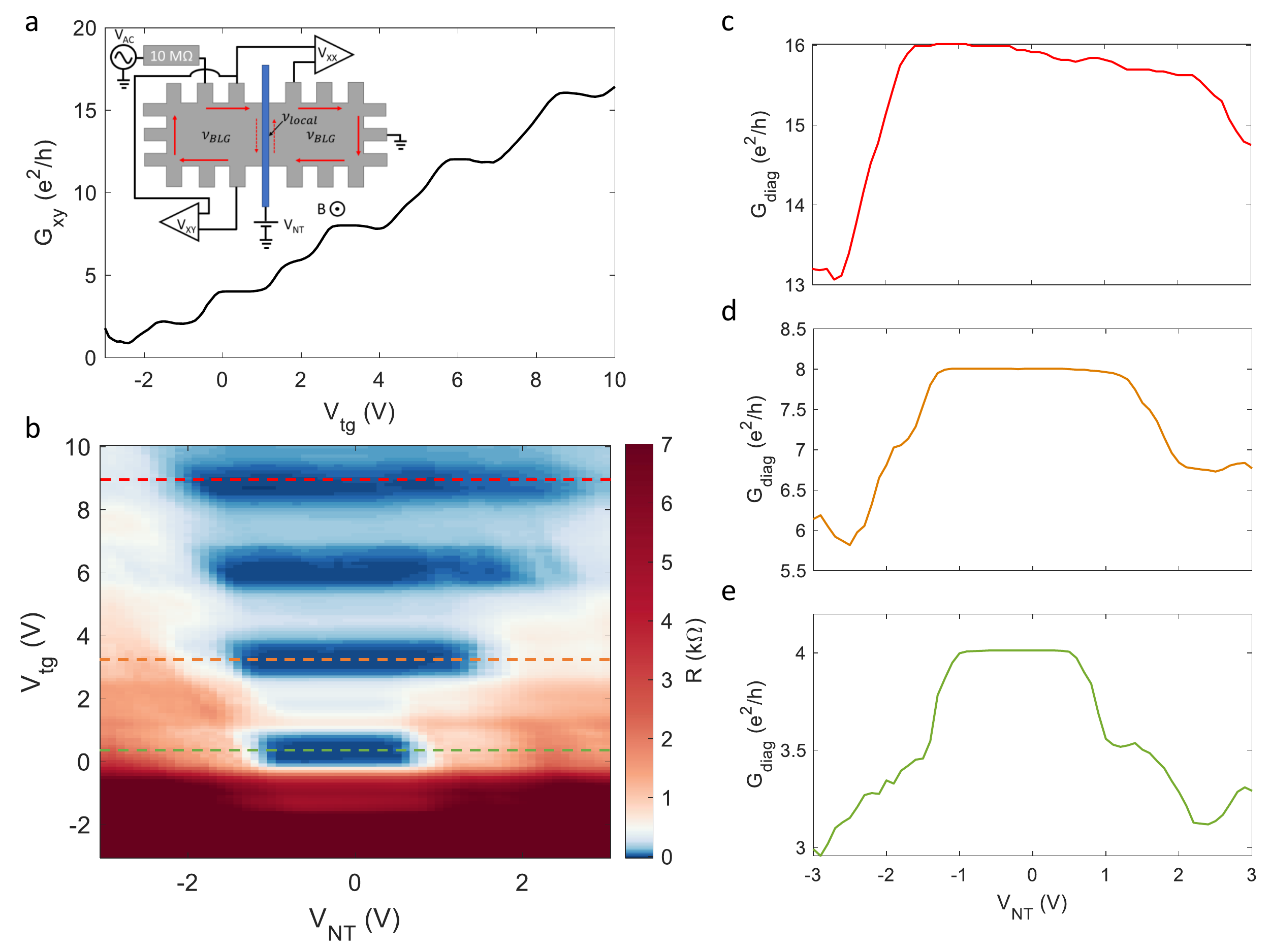}
    \caption{SWNT-BLG device in the quantum Hall regime, at $T = 1.6$ K and $V_{bg}=0$ V. (a) Hall conductance $G_{xy}$ of bulk channel as a function of top gate voltage $V_{tg}$. SWNT local gate voltage $V_{NT}$ does not affect the value. Inset: schematic of measurement circuit. (b) $R_{xx}$ of bulk channel across SWNT local gate as a function of $V_{tg} and V_{NT}$. (c)-(e) Line cuts of $G_{diag}=(R_{xx}+R_{xy})^{-1}=\nu_{local} (e^2/h)$ at fixed $V_{tg}$ (indicated by dashed lines on (b)) corresponding to integer $\nu_{bulk}$. (c) $\nu_{bulk}=16$. (d) $\nu_{bulk}=8$. (e) $\nu_{BLG}=4$.}
    \label{fig:CNTgating}
\end{figure}

\begin{figure}
    \centering
    \includegraphics[width=0.9\textwidth]{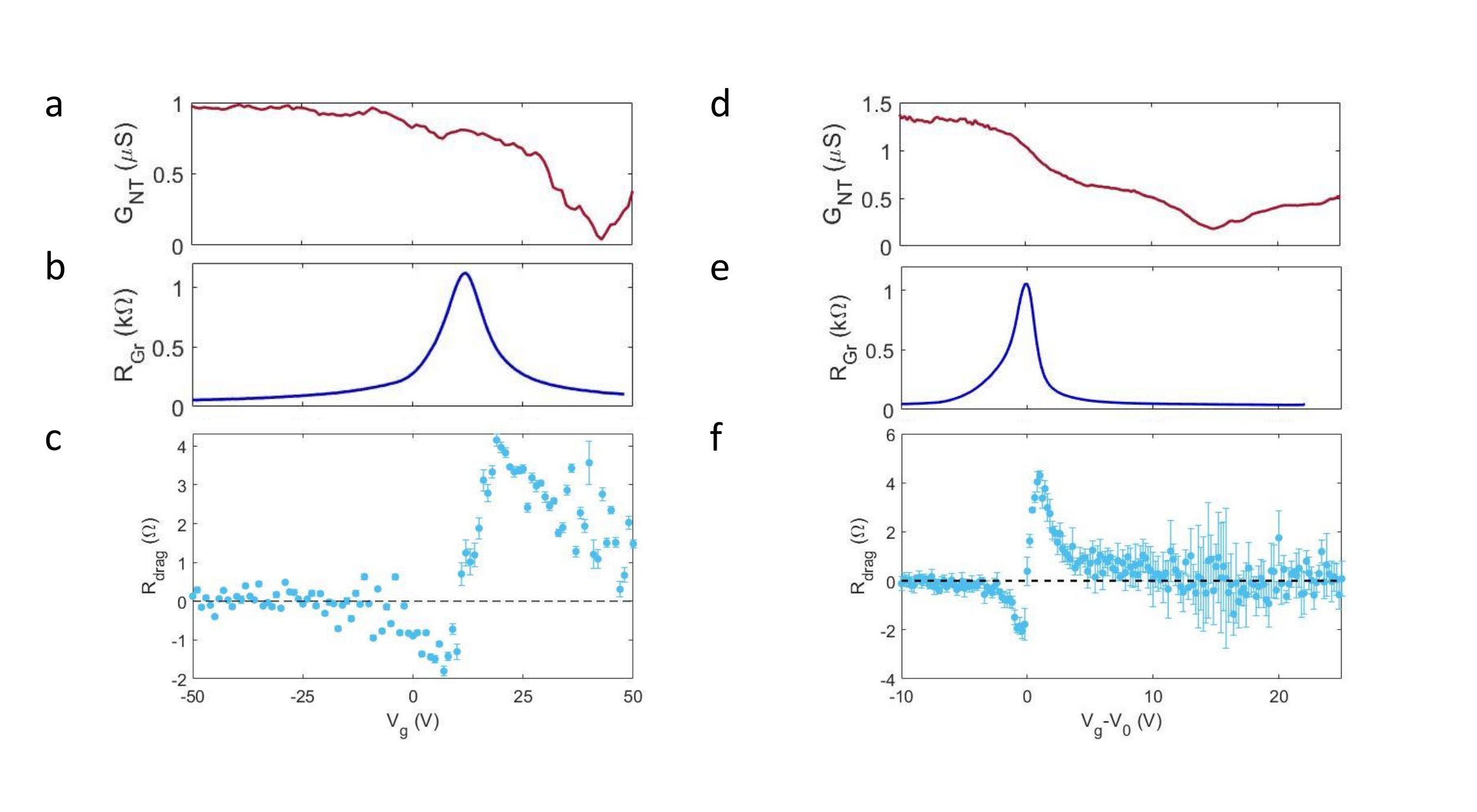}
    \caption{Individual layer transport and drag resistance as a function of back gate voltage for measurement D1-A. (a) SWNT conductance at $T = 200\; \textnormal{K}$. (b) Graphene resistance at $T = 100 \;\textnormal{K}$. (c) Drag resistance at $T = 200$ K. (d-e) Individual layer transport and drag resistance as a function of adjusted back gate voltage for the same device after thermal cycling and 9 months of storage (measurement D1-B). The voltage value at the graphene charge neutrality point has been subtracted. (d) SWNT conductance at $T = 200$ K. (e) Graphene resistance at $T = 200$ K. (f) Drag resistance at $T = 200$ K.}
    \label{fig:S8}
\end{figure}

\begin{figure}
    \centering
    \includegraphics[width=0.9\textwidth]{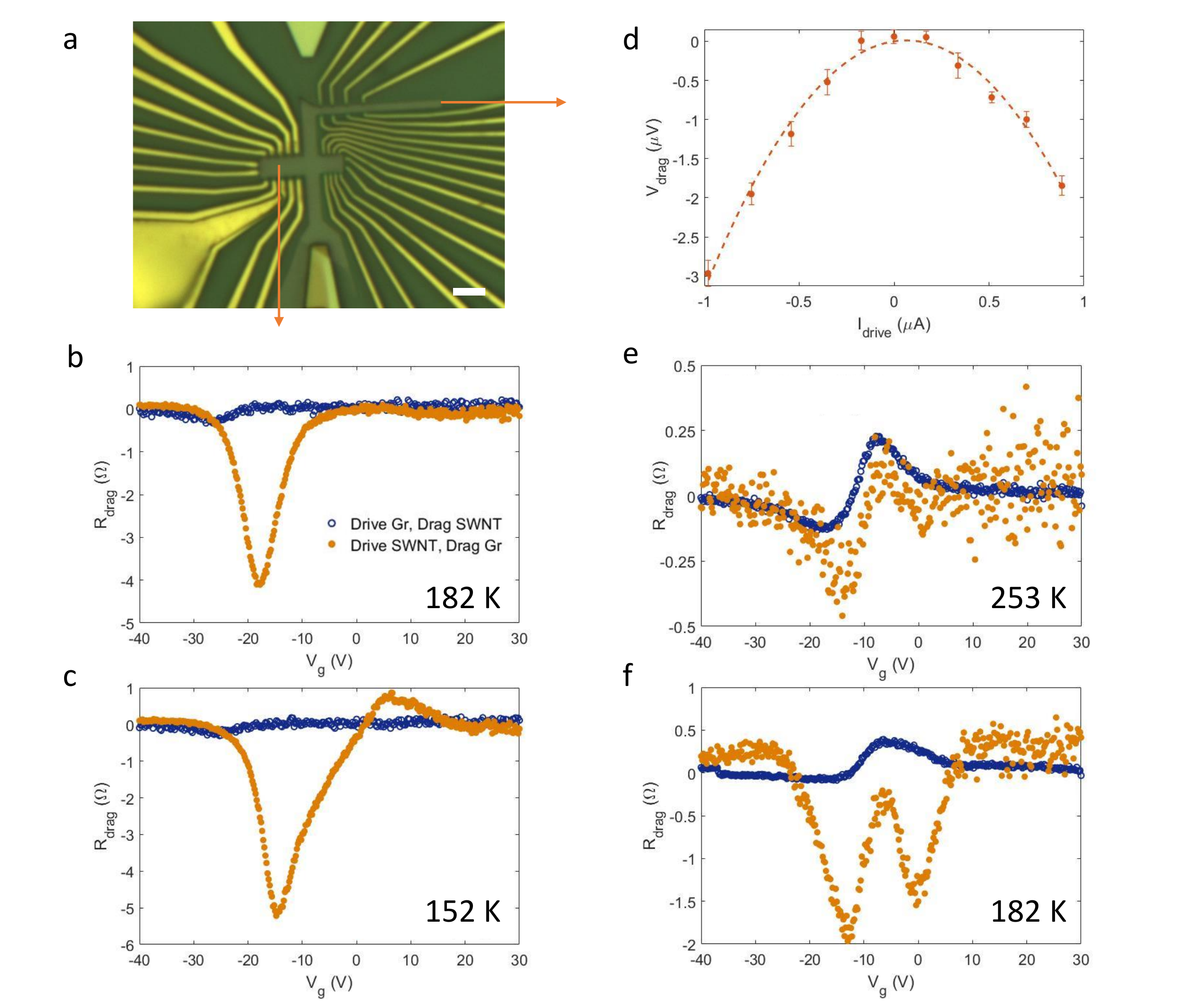}
    \caption{a) Optical microscope image of device D2. D2-1 refers to the lower, wider section and D2-2 is the upper, narrower section. Scale bar is 4 $\mu$m. (b-c) D2-1 drag resistance versus back gate voltage using graphene (blue) or CNT (orange) as the drive layer, at T = 182 K (b) and 152 K (c). (d) D2-2 drag voltage in graphene versus drive current in SWNT at $T =$ 200 K. (e-f) D2-2 drag resistance versus back gate voltage using graphene (blue) or SWNT (orange) as the drive layer, at $T =$ 253 K (e) and 182 K (f). The CNT charge neutrality point in all measurements is at $\sim 40\;\textnormal{V}$.}
    \label{fig:S9}
\end{figure}

\begin{figure}
    \centering
    \includegraphics[width=0.9\textwidth]{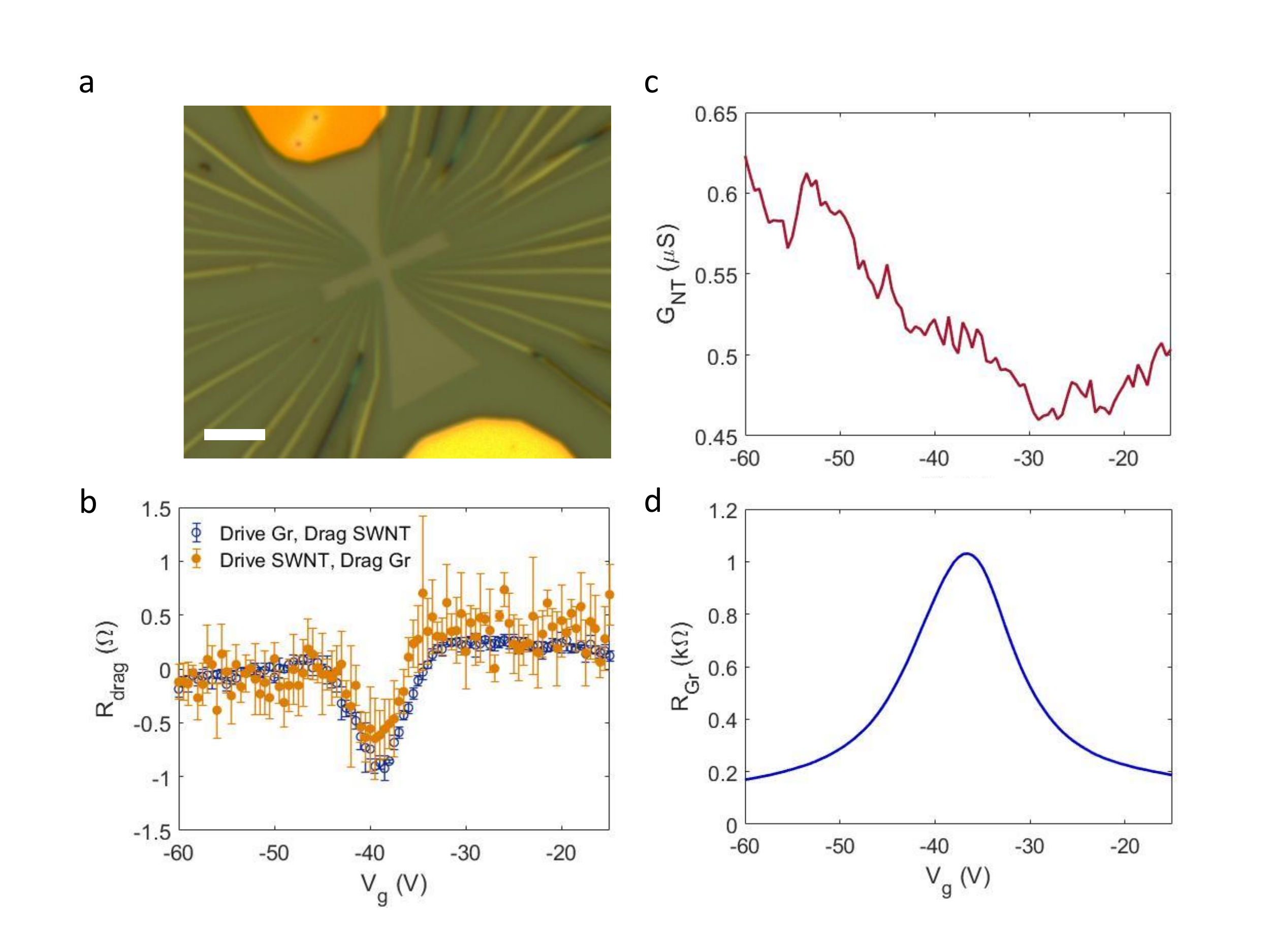}
    \caption{a) Optical microscope image of device D3. Scale bar is 5 $\mu$m. (b) D3 drag resistance versus gate voltage using graphene (blue) or SWNT (orange) as the drive layer, at $T$ = 200 K. (c) SWNT conductance versus gate voltage at $T$ = 200 K. (d) Graphene conductance versus gate voltage at $T$ = 200 K.}
    \label{fig:S10}
\end{figure}

\clearpage

%\bibliographystyleS{apsrev4-2}
%\bibliographyS{CNT-Gr-SM-v3}% Produces the bibliography via BibTeX.


\end{document}